\documentclass[aps,prb,twocolumn,amsmath,amssymb,revsymb,letter]{revtex4-1}
\usepackage{graphicx}
\usepackage{color} 
\usepackage{dcolumn}
\usepackage{bm}
\usepackage{setspace}
\usepackage{sidecap}
\usepackage{natbib}
\usepackage{multirow}
\usepackage{hyperref}

\providecommand{\e}{\varepsilon}
\providecommand{\h}{\hbar}
\providecommand{\p}{\partial}
\newcommand{\kk}{\mathbf{k}}
\newcommand{\rr}{\mathbf{r}}
\newcommand{\pp}{\mathbf{p}}
\newcommand{\RR}{\mathbf{R}}
\newcommand{\Ruc}{\boldsymbol{\mathcal{R}}}
\newcommand{\Ss}{\mathbf{S}}
\newcommand{\Jj}{\mathbf{J}}
\newcommand{\Ll}{\mathbf{L}}
\newcommand{\bfd}{\mathbf{d}}
\newcommand{\bfrho}{\boldsymbol{\rho}}
\newcommand{\II}{\mathbf{I}}
\newcommand{\Te}{\scriptsize{\textrm{Te}}}
\newcommand{\Hg}{\scriptsize{\textrm{Hg}}}
\newcommand{\op}{\uparrow}
\newcommand{\ned}{\downarrow}

\begin{document}

\title{Hyperfine interactions in two-dimensional HgTe topological insulators}

\author{Anders Mathias Lunde$^{1,2,3}$ and Gloria Platero$^{1}$}
\affiliation{$^1$Instituto de Ciencia de Materiales de Madrid (ICMM),  
Consejo Superior de Investigaciones Cient\'ificas (CSIC),
28049 Madrid, Spain\\
$^2$Instituto de Estructura de la Materia, CSIC, Serrano 123, 28006 Madrid, Spain\\
$^3$Center for Quantum Devices, Niels Bohr Institute,
University of Copenhagen, 
Denmark}
\date{\today}

\begin{abstract}
We study the hyperfine interaction between the nuclear spins and the electrons in a HgTe quantum well, which is the prime experimentally realized example of a two-dimensional topological insulator. The hyperfine interaction is a naturally present, internal source of broken time-reversal symmetry from the point of view of the electrons. The HgTe quantum well is described by the so-called Bernevig-Hughes-Zhang (BHZ) model. The basis states of the BHZ model are combinations of both $S$- and $P$-like symmetry states, which means that three kinds of hyperfine interactions play a role: (i) The Fermi contact interaction, (ii) the dipole-dipole-like coupling and (iii) the electron-orbital to nuclear-spin coupling. We provide benchmark results for the forms and magnitudes of these hyperfine interactions within the BHZ model, which give a good starting point for evaluating hyperfine interactions in \emph{any} HgTe nanostructure. We apply our results to the helical edge states of a HgTe two-dimensional topological insulator and show how their total hyperfine interaction becomes anisotropic and dependent on the orientation of the sample edge within the plane. Moreover, for the helical edge states the hyperfine interactions due to the $P$-like states can dominate over the $S$-like contribution in certain circumstances. 
\end{abstract}

\maketitle

\section{Introduction}

A topological insulator (TI) host gapless surface or edge states, while the bulk of the material has an insulating energy gap.\cite{Qi-Zhang-review-RMP-2010,Kane-Mele-first-paper-PRL-2005,Kane-Mele-second-paper-PRL-2005,Hasan-review-RMP-2010} In three-dimensional TIs the gapless surface states are spin-polarized two-dimensional (2D) Dirac fermions, whereas 2D TIs contain one-dimensional (1D) helical edge states. The helical edge states appear in counterpropagating pairs, and the states with equal energy and opposite wave numbers, $k$ and $-k$, form a Kramers pair. Thus, elastic scattering from one helical edge state (HES) to the other one within a pair cannot be induced by time-reversal invariant potentials e.g.~stemming from impurities.\cite{Xu-Moore-PRB-2006} Therefore, the transport through a 2D TI is to a large extend ballistic with a quantized conductance of $e^2/h$ per pair of HESs.  This highlights the central role of time-reversal symmetry in TIs. 

Quantized conductance have recently been measured in micron-sized samples in HgTe quantum wells,\cite{Konig-Science-2007,Roth-Science-2009,Konig-JPSJ-review-2008,Buhmann-JAP-2011,Brune-Molenkamp-nature-phys-2012,Konig-et-al-PRX-2013} which to date is the most important experimental demonstration of a 2D TI. Evidence of edge state transport was found in both two-terminal\cite{Konig-Science-2007} and multi-terminal\cite{Roth-Science-2009} devices. Moreover, clever experiments combining the metallic spin Hall effect and a 2D TI in a HgTe quantum well (QW) demonstrated the connection between the spin and the propagation direction.\cite{Brune-Molenkamp-nature-phys-2012} However, also deviations from perfect conductance have been observed in longer HgTe devices,\cite{Konig-Science-2007,Roth-Science-2009,Konig-et-al-PRX-2013,Gusev-et-al-PRB-2011} which could stem from e.g.~inelastic scattering mechanisms.\cite{Schmidt-Rachel-Oppen-Glazman-PRL-2012,Budich-Dolcini-Recher-Trauzettel-PRL-2012,Lezmy-Oreg-Berkooz-PRB-2012,Crepin-et-al-PRB-2012,Maciejko-PRL-2009,Strom-Johannesson-Japaridze-PRL-2011} The effect of external magnetic fields have also been considered.\cite{Konig-Science-2007,Konig-JPSJ-review-2008,Hankiewicz-PRL-2010,Maciejko-Qi-Zhang-PRB-2010,Scharf-Matos-Abiague-Fabian-PRB-2012,Delplace-Li-Buttiker-PRL-2012,Kharitonov-PRB-2012,Gusev-et-al-PRL-2013} The TI state in HgTe QWs was predicted by Bernevig, Hughes and Zhang (BHZ)\cite{Bernevig-Zhang-Science-2006} by using a simplified $\kk\cdot\pp$ model containing states with $S$- and $P$-like symmetry, respectively. They found that beyond a critical thickness of the HgTe QW, the TI state would appear as confirmed experimentally.\cite{Konig-Science-2007,Roth-Science-2009,Konig-JPSJ-review-2008,Konig-JPSJ-review-2008,Buhmann-JAP-2011} Furthermore, interesting experimental progress on 2D TI properties has also be achieved in InAs/GaSb QWs\cite{Knez-Du-Sullivan-PRL-2011,Suzuki-et-al-PRB-2013,Du-Knez-et-al-arxiv-2013} as proposed theoretically.\cite{Liu-Hughes-Zhang-et-al-PRL-2008}

\begin{figure}
\includegraphics[width=0.32\textwidth,angle=0]{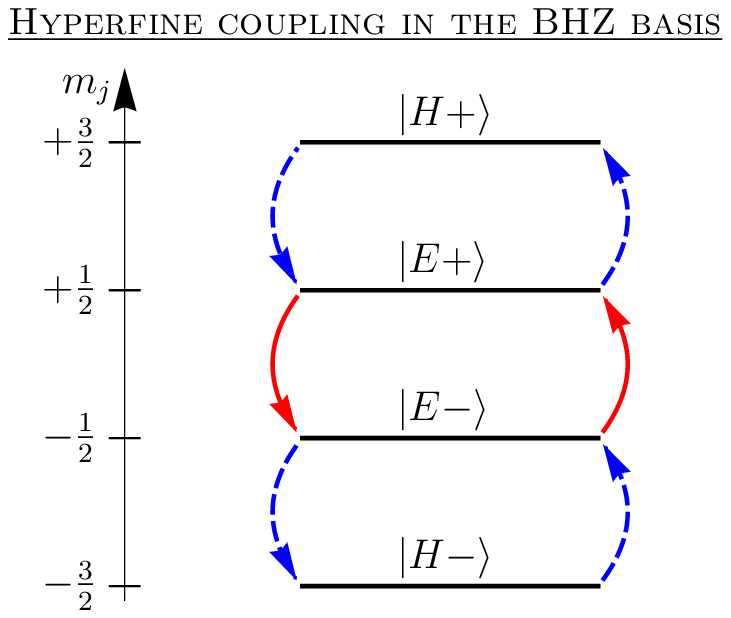} 
\caption{Hyperfine coupling between the BHZ basis states $\{ |H+\rangle$, $|E+\rangle$, $|E-\rangle$, $|H-\rangle \}$, which have the total angular momentum projections on the $z$-axis $m_j$ as indicated. Every increase (decrease) of $m_j$ in the electronic sector is accompanied by a decrease (increase) of a nuclear spin due to angular momentum conservation. Hyperfine interactions due to both $S$- and $P$-like states connect the two time-reversed blocks of the BHZ model (red full arrows). However, \emph{only} hyperfine interactions due to $P$-like states connect states within a single time-reversed block (blue dashed arrows).}
\label{fig:Spin-flips-between-BHZ-states}
\end{figure}

Hyperfine (HF) interactions between the electron and nuclear spins can play an important role in nanostructures -- even though it is often weak.\cite{Coish-Baugh-review-2009,Schliemann-Khaetskii-Loss-JoPCM-Review-2003,Slichter-BOOK,Stoneham-BOOK}  For instance in quantum dots, HF interactions can limit the coherence of single electronic spins\cite{Khaetskii-Loss-Glazman-PRL-2002,Koppens-Nowack-Vandersypen-PRL-2008,Petta-et-al-Science-2005,Cywinski-decoherence-QD-review-2011} and, moreover, it can even lead to current hysteresis due to bistability of the dynamical nuclear spin polarization.\cite{Ono-Tarucha-PRL-2004,Pfund-Shorubalko-Ensslin-Leturcq-PRL-2007,Rudner-Levitov-PRL-2007,Lunde-et-al-DNP-in-DQD-2013} HF-induced nuclear spin ordering in interacting 1D\cite{Braunecker-Simon-Loss-PRL-2009,Braunecker-Simon-Loss-PRB-2009,Scheller-et-al-arxiv-2013} and 2D\cite{Simon-Loss-PRL-2007,Braunecker-Simon-Loss-PRB-2008} systems have also been discussed. Most studies consider the so-called  contact HF interaction,\cite{Coish-Baugh-review-2009,Schliemann-Khaetskii-Loss-JoPCM-Review-2003,Slichter-BOOK,Stoneham-BOOK} which is relevant for electrons in orbital states with $S$-like symmetry e.g.~the conduction band in GaAs. However, for $P$-like orbital states -- such as the valence band in GaAs -- the contact HF interaction is absent. Nevertheless, other anisotropic HF interactions are present for $P$-like states such as the dipole-dipole-like HF interaction,\cite{Fischer-Coish-Bulaev-Loss-PRB-2008,Fischer-Loss-PRL-2010,Fischer-Trauzettel-Loss-PRB-2009,Fischer-Trif-Coish-Loss-Solid-state-comm-2009,Testelin-Bernardot-Eble-Chamarro-PRB-2009} which can play a significant role e.g. for the decoherence of a hole confined in a quantum dot.\cite{Fischer-Coish-Bulaev-Loss-PRB-2008,Fischer-Loss-PRL-2010,Testelin-Bernardot-Eble-Chamarro-PRB-2009,Chekhovich-et-al-nat-phys-2013}

HF interactions and dynamical nuclear spin polarization have also been investigated in the context of integer quantum Hall systems,\cite{Dobers-Klitzing-et-al-PRL-1988,Wald-Kouwenhoven-McEuen-et-al-PRL-1994,Kim-Vagner-Xing-PRB-1994,Dixon-McEuen-et-al-PRB-1997,Deviatov-et-al-PRB-2004,Wurtz-et-al-PRL-2005,Nakajima-Kobayashi-Komiyama-PRB-2010,Nakajima-Komiyama-PRB-2012} which contain unidirectional edge states. Here HF-induced spin-flip transitions between the unidirectional edge states can create nuclear spin polarization locally at the boundary of the 2D sample.\cite{Wald-Kouwenhoven-McEuen-et-al-PRL-1994,Kim-Vagner-Xing-PRB-1994,Dixon-McEuen-et-al-PRB-1997,Deviatov-et-al-PRB-2004,Wurtz-et-al-PRL-2005,Nakajima-Kobayashi-Komiyama-PRB-2010,Nakajima-Komiyama-PRB-2012} Recently, we have predicted a similar phenomenon for a 2D TI, namely that embedded fixed spins such as the nuclear spins in a 2D TI can polarize locally at the boundary due to a current through the HESs.\cite{Lunde-Platero-PRB-2012} Interestingly, the 2D TI with localized spins remains ballistic,\cite{Tanaka-Furusaki-Matveev-PRL-2011,Lunde-Platero-PRB-2012} except if additional spin-flip mechanisms for the localized spins are present.\cite{Lunde-Platero-PRB-2012} However, combining localized spins and Rashba spin-orbit coupling in the 2D TI can produce a conductance change.\cite{Eriksson-Strom-Sharma-Johannesson-PRB-2012,Maestro-Hyart-Rosenow-PRB-2013,Eriksson-PRB-2013} In the previous works,\cite{Lunde-Platero-PRB-2012,Tanaka-Furusaki-Matveev-PRL-2011,Eriksson-Strom-Sharma-Johannesson-PRB-2012,Maestro-Hyart-Rosenow-PRB-2013,Eriksson-PRB-2013} the interaction between the fixed spins embedded into the 2D TI and the HESs were modelled phenomenologically. In contrast, here we pay special attention to the detailed forms of the HF interactions within a 2D TI. 

In this paper, we find the different HF interactions within the BHZ model for a HgTe QW. To this end, we take into account both the $S$- and $P$-like states of the BHZ model, which couple differently to the nuclear spins. We show that all the HF Hamiltonians couple the time-reversed blocks of the BHZ model. However, only HF interactions relevant for $P$-like states couple states \emph{within} a time-reversed block as illustrated in Fig.\ref{fig:Spin-flips-between-BHZ-states}. Moreover, we estimate the different HF coupling constants. The derived Hamiltonians are general in the sense that they can be used to find the HF interactions for any kind of nanostructure in a HgTe QW, e.g. quantum dots,\cite{Chang-Lou-PRL2011} ring structures,\cite{Michetti-Recher-2011} quantum point contacts\cite{Zhang-et-al-PRB-2011} or hole structures.\cite{Shan-et-al-PRB-2011} As an illustrative example, we find the HF interactions for a pair of HESs. Remarkably, the intra HES transitions coupled to \emph{all} the nuclear spin components perpendicular to the  propagation direction of the HESs. This kind of coupling is unusual compared to e.g. an ordinary Heisenberg model. Interestingly, the details of the HF interactions depend on the spacial direction of the boundary at which the HESs propagate.  
  
The paper is structured as follows: First the HF interactions and the BHZ model are outlined in secs. \ref{sec:HF-int} and \ref{sec:BHZ-model}. Then the HF interactions are found within the BHZ model for the simplest case of a 2D QW (sec.\ref{sec:HF-in-BHZ}). From this, we derive the HF interactions for a given nanostructure in sec. \ref{sec:HF-any-nanostruc}. Finally, the HF interactions for the HESs are found and discussed (sec. \ref{sec:HF-for-HES}). Appendices \ref{App:TR-af-basis-states}-\ref{app:HF-in-spin-operators} provide various details for completeness.

\section{The Hyperfine Interactions}\label{sec:HF-int}

The HF interaction between an electron at position $\rr$ with spin $\Ss=(S_x,S_y,S_z)$ and the nuclear spin $\II_n=(I_{x,n},I_{y,n},I_{z,n})$ of the lattice atom at $\RR_n$ can be derived from the Dirac equation\cite{Stoneham-BOOK,Fischer-Trif-Coish-Loss-Solid-state-comm-2009} to be (in SI units)
\begin{subequations}
\label{eq:HF123-of-nuclei-nr-n}
\begin{align}
h^n_1&=\frac{\mu_0}{4\pi}\frac{8\pi}{3} 
\gamma_e\gamma_{j_n} \delta(\rr_n)\Ss\cdot\II_n,
\label{eq:HF1-of-nuclei-nr-n}\\
h^n_2&=\frac{\mu_0}{4\pi}\gamma_e\gamma_{j_n}
\frac{3(\mathbf{e}_n\cdot\Ss)(\mathbf{e}_n\cdot\II_n)-\Ss\cdot\II_n}{r_n^3(1+r_c/r_n)},
\label{eq:HF2-of-nuclei-nr-n}\\
h^n_3&=\frac{\mu_0}{4\pi}\gamma_e\gamma_{j_n}
\frac{\mathbf{L}_n\cdot\II_n}{r_n^3(1+r_c/r_n)},
\label{eq:HF3-of-nuclei-nr-n}
\end{align}
\end{subequations}
where $h_1^n$ is the Fermi contact interaction\cite{Fermi-original-1930}, $h_2^n$ is the dipole-dipole like coupling between the electrons spin and the nuclear spin, and $h_3^n$ is the coupling of the electrons orbital momentum $\mathbf{L}_n=\rr_n\times \pp$ and the nuclear spin. Here $\rr_n=\rr-\RR_n$ is the electrons position relative to the $n$th nucleus, $r_n\equiv|\rr_n|$, $\mathbf{e}_n\equiv\rr_n/r_n$ and $\mu_0$ is the vacuum permeability. The gyromagnetic ratios of the electron $\gamma_e$ and the $n$th nuclear spin $\gamma_{j_n}$ of the isotope $j$ are, respectively, given by ($e>0$) $\gamma_e=g_e\mu_B/\hbar$, where $\mu_B=e\hbar/(2m_e)$ is the Bohr magneton and $g_e\simeq2$ the electron g factor; and $\gamma_{j_n}=g_{j_n}\mu_N/\hbar$, where $\mu_N=e\hbar/(2m_p)=\mu_B/1836$ is the nuclear magneton and $g_{j_n}$ the g factor of the $j$th isotope. Here $m_e$ and $m_p$ are the bare electron and proton mass, respectively.\cite{footnote-gyromagnetic-ratios-def} Moreover, $r_c$ is a length scale related to the finite size of the nucleus and therefore much smaller than all other length scales in the system. It can be found to be\cite{Stoneham-BOOK} $r_c=Ze^2/(2mc^2)\simeq Z\times 1.5$fm, where $Z$ is the number of protons in the nucleus.\cite{footnote-r_c-cut-off} Thus, the total HF interaction between an electron and all the nuclear spins in the lattice is
\begin{align}
H_{\textrm{HF}}&=
H_{\textrm{HF},1}+H_{\textrm{HF},2}+H_{\textrm{HF},3}
\nonumber\\
&=\sum_{n}h^n_1+\sum_{n}h^n_2+\sum_{n}h^n_3,
\label{eq:H-HF-all}
\end{align}
where only those lattice points $\RR_n$ with a non-zero nuclear spin are included in the sum. 

Not every atom in a HgTe crystal has a non-zero nuclear spin in contrast to e.g. GaAs. The amount of stable isotopes with a non-zero spin in Hg and Te are about\cite{Schliemann-Khaetskii-Loss-JoPCM-Review-2003} 
\begin{align}
17\%
\ \textrm{of}\ ^{199}\textrm{Hg}\ 
(\textrm{spin-}1/2),& 
\quad 
\! 13\%
\ \textrm{of}\ ^{201}\textrm{Hg}\ 
(\textrm{spin-}3/2), 
\nonumber\\
1\%
\ \textrm{of}\ ^{123}\textrm{Te}\ 
(\textrm{spin-}1/2),& 
\quad
\ 7\%
\ \textrm{of}\ ^{125}\textrm{Te}\ 
(\textrm{spin-}1/2). 
\label{eq:isotopes-of-HgTe}
\end{align}
Hence, about 19$\%$ of all the atoms in HgTe have a non-zero nuclear spin. By isotope selection processes, this number can be varied somewhat experimentally. 

The contact  interaction $H_{\textrm{HF},1}$ is the only important HF interaction for $S$-like states due to their spherical symmetry around the atomic core. On the other hand, $P$-like states vanish at the atomic core and therefore the contact interaction does not affect electrons in those states. In contrast, the two other terms $H_{\textrm{HF},2}$ and $H_{\textrm{HF},3}$ can indeed play a role for $P$-like states such as heavy-holes.\cite{Fischer-Coish-Bulaev-Loss-PRB-2008} Moreover, Fischer \emph{et al.}\cite{Fischer-Coish-Bulaev-Loss-PRB-2008} found the atomic HF coupling constants to be about one order of magnitude lower for $P$-like compared to $S$-like states in GaAs.

\section{The Bernevig-Hughes-Zhang (BHZ) model}\label{sec:BHZ-model}

Bernevig, Hughes and Zhang\cite{Bernevig-Zhang-Science-2006} constructed a simple model describing the basic physics of a HgTe QW. The effective  $4\times4$ BHZ Hamiltonian is derived using $\kk\cdot\pp$ methods\cite{Winkler-BOOK-2003,Bastard-BOOK,Fabian-review-2007} and valid for $\kk=(k_x,k_y)$ close to the $\Gamma$ point, i.e.~close to $\kk=(0,0)$. The basis states of the model are the two Kramer pairs $|E\pm\rangle$ and $|H\pm\rangle$. Details on the derivation of the BHZ model are found in Refs.~\onlinecite{Bernevig-Zhang-Science-2006,Qi-Zhang-review-RMP-2010,Hankiewicz-NJP-2010}. For a 2D QW the BHZ Hamiltonian is 
\begin{subequations}
\label{eq:H-BHZ-model}
\begin{align}
\mathcal{H}_0= 
\sum_{{\kk}} \mathbf{c}_{\kk}^\dag H_0(\kk)\mathbf{c}_{\kk}^{},
\label{eq:H-BHZ-model-periodic-boundary}
\end{align}
where $\mathbf{c}_{\kk}^\dag=(c_{\kk,E+}^\dag,c_{\kk,H+}^\dag,c_{\kk,E-}^\dag,c_{\kk,H-}^\dag)$ is a vector of creation operators and      
\begin{align}
H_0(\kk)=
\left(
\begin{array}{cc}
  h(\kk) & \mathbf{0} \\
  \mathbf{0} & h^\ast(-\kk)
\end{array}
\right)
\label{eq:H-BHZ-model-matrix1}
\end{align}
with $\mathbf{0}$ being a zero $2\times2$ matrix and
\begin{align}
h(\kk)=
\left(
\begin{array}{cc}
 \e_k+M_k & A(k_x+ik_y) \\
 A(k_x-ik_y) & \e_k-M_k \\
\end{array}
\right).
\label{eq:H-BHZ-model-matrix2}
\end{align}
\end{subequations}
Here $\e_k=-Dk^2$, $M_k=M_0-Bk^2$, and $k\equiv \sqrt{k_x^2+k_y^2}$ have been introduced.\cite{footnote-zero-energy} The parameters $A$, $B$, $D$ and $M_0$ depend on the QW geometry.\cite{Bernevig-Zhang-Science-2006,Qi-Zhang-review-RMP-2010} Importantly, varying the QW width changes the sign of $M_0$, which in turn makes the system go from a non-topological to a topological state with HESs.\cite{Bernevig-Zhang-Science-2006}   

The hamiltonian (\ref{eq:H-BHZ-model-periodic-boundary}) a priori has periodic boundary conditions and thereby does not contain any edges. By introducing boundaries into the model, it is possible to derive explicitly the HESs in the TI state of the QW.\cite{Zhou-edge-states-PRL-2008,Bihlmayer-Edge-states-in-Bi-films-PRB-2011} This will be discussed further in sec.~\ref{sec:HES}. 

Within the envelope function approximation\cite{Winkler-BOOK-2003,Bastard-BOOK,Fabian-review-2007} the states of the BHZ model are
\begin{subequations}
\label{eq:basis-states-mellem2}
\begin{align}
|E+\rangle&=f_{E\Gamma_6}(z)|\Gamma_6,+1/2\rangle+f_{E\Gamma_8}(z)|\Gamma_8,+1/2\rangle,\\   
|H+\rangle&=f_{H}(z)|\Gamma_8,+3/2\rangle,\\   
|E-\rangle&=f_{E\Gamma_6}(z)|\Gamma_6,-1/2\rangle+f_{E\Gamma_8}(z)|\Gamma_8,-1/2\rangle,\\   
|H-\rangle&=f_{H}(z)|\Gamma_8,-3/2\rangle,
\end{align}
\end{subequations}
where $f_i(z)$ are the transverse envelope functions in the $z$-direction perpendicular to the 2D QW and $|\Gamma_i, m_j\rangle$ are the lattice periodic functions\cite{footnote-lattice-periodic-func}  at $\kk=0$ for the $\Gamma_i$ band with projection $m_j$ of the total angular momentum, $\Jj=\Ll+\Ss$, on the $z$-axis. Here $\Ss$ is the electron spin and $\Ll$ is the orbital angular momentum (see Appendix \ref{App:TR-af-basis-states}). The time-reversal operator $\Theta$ connects states within a Kramer pair ($\Theta|E\pm\rangle=\mp|E\mp\rangle$ and $\Theta|H\pm\rangle=\mp|H\mp\rangle$), and the two blocks in $H_0(\kk)$ (\ref{eq:H-BHZ-model-matrix1}) are related by time-reversal. Here we choose phase conventions of the envelope functions such that time-reversed partners have equal envelope functions. Moreover, $f_{E\Gamma_6}$ and $f_H$ are chosen real, whereas $f_{E\Gamma_8}$ is chosen purely imaginary. (Appendix \ref{App:TR-af-basis-states} gives more details on the envelope functions and the lattice periodic functions.)

The states $|E\pm\rangle$ are seen to be mixtures of the $S$-like $\Gamma_6$ band and the $P$-like $\Gamma_8$ band with $m_j=\pm1/2$, whereas $|H\pm\rangle$ consist only of the $P$-like $\Gamma_8$ band with $m_j=\pm3/2$. Hence, the states have a definite total angular momentum projection,  
\begin{align}
J_z|E\pm\rangle=\pm\frac{1}{2}\h|E\pm\rangle
\quad\textrm{and}\quad
J_z|H\pm\rangle=\pm\frac{3}{2}\h|H\pm\rangle,
\label{eq:mz-of-BHZ-states}
\end{align}  
but $|E\pm\rangle$ are not  eigenstates of $\Jj^2$. 

The HF interactions can only induce transitions between states with a difference of angular momentum projection of one unit: $m_{j}-m_{j'}=\pm1$. Therefore, we can already at this point see that only particular combinations of the BHZ states can be connected by HF interactions as seen in Fig.~\ref{fig:Spin-flips-between-BHZ-states}. Furthermore, it is evident that HF interactions relevant for both $S$- and $P$-like states need to be included to have a full description of the HF interactions in a HgTe TI.

The real-space basis functions of $\mathcal{H}_0$ (\ref{eq:H-BHZ-model-periodic-boundary}) for the 2D QW with periodic boundary conditions are 
\begin{subequations}
\label{eq:basis-for-BHZ-2D}
\begin{align}
\varphi^{}_{\kk,E\pm}(\rr)
=&\frac{\sqrt{v_a}}{\sqrt{L_xL_y}}e^{i(k_xx+k_yy)}
\\
&{\times}\Big[f_{E\Gamma_6}^{}(z)u^{}_{\Gamma_6,\pm\frac{1}{2}}(\rr)
+f^{}_{E\Gamma_8}(z)u_{\Gamma_8,\pm\frac{1}{2}}^{}(\rr)\Big],\nonumber\\   
\varphi^{}_{\kk,H\pm}(\rr)
=&\frac{\sqrt{v_a}}{\sqrt{L_xL_y}}e^{i(k_xx+k_yy)} 
f_H^{}(z)u^{}_{\Gamma_8,\pm\frac{3}{2}}(\rr),
\label{eq:basis-for-BHZ-2D-H+}
\end{align}
\end{subequations}
where $\rr=(x,y,z)$, $L_x$ ($L_y$) is the QW length in the $x$ ($y$) direction and $u_{\Gamma_i,m_j}(\rr)\equiv\langle\rr|\Gamma_i,m_j\rangle$ are the real-space lattice periodic functions at $\kk=0$. Moreover, we have included the atomic volume\cite{footnote-atomic-vol} $v_a$ explicitly here as it is often done for HF related calculations.\cite{Fischer-Loss-PRL-2010,Fischer-Trauzettel-Loss-PRB-2009,Fischer-Trif-Coish-Loss-Solid-state-comm-2009,Fischer-Coish-Bulaev-Loss-PRB-2008} It depends on the choice of the individual normalization of the envelope functions and the lattice periodic functions, respectively, if $v_a$ should be included explicitly,\cite{Coish-Baugh-review-2009} as discussed in Appendix \ref{App:normalization}.

\section{Hyperfine interactions within the BHZ model}\label{sec:HF-in-BHZ}

Next, we find the HF interactions within the BHZ model by using the states (\ref{eq:basis-for-BHZ-2D}) for a 2D QW with periodic boundary conditions. As we shall see, these results are useful, since they allow us to find the HF interactions for any nanostructure created in a HgTe QW (sec. \ref{sec:HF-any-nanostruc}).

\subsection{Outline of the way to find the hyperfine interaction matrix elements}\label{subsec:general-way-to-find-the-matrix-element}

The HF interactions (\ref{eq:HF123-of-nuclei-nr-n}) are local in space on the atomic scale, so the important part of the wavefunction with respect to the HF interactions is the behavior around the nucleus. Hence, in the envelope function approximation, it is the rapidly varying lattice periodic functions $u_{\Gamma_i,m_j}(\rr)$ that play the central role, whereas the slowly varying envelope functions only are multiplicative factors at the atomic nucleus, as we shall see below.

We set out to find the HF interactions 
\begin{align}
\mathcal{H}_{\textrm{HF},i}= 
\sum_{\kk,\kk'}
\sum_{\substack{\upsilon,\upsilon'=E,H \\ \tau\tau'=\pm}} 
\!\!\!\langle\varphi_{\kk \upsilon\tau}|H_{\textrm{HF},i}|\varphi_{\kk' \upsilon'\tau'}\rangle 
c_{\kk \upsilon\tau}^\dag c_{\kk' \upsilon'\tau'}^{}
\label{eq:HF-H-2nd-quantized-general}
\end{align}
for $i=1,2,3$ in the basis (\ref{eq:basis-for-BHZ-2D}), i.e.~for $\varphi_{\kk,\upsilon\pm}(\rr)$ with $\upsilon=E,H$. We begin by describing the general way that we find the HF interaction matrix elements $\langle\varphi_{\kk \upsilon\tau}|H_{\textrm{HF},i}|\varphi_{\kk' \upsilon'\tau'}\rangle$. To this end, the integration over the entire system volume $\mathcal{V}$ is rewritten as  a sum of integrals over each unit cell $m$ of volume $v_{uc}^{(m)}$, i.e.
\begin{align}
\int_{\mathcal{V}}d \rr (\cdots)
=\sum_{\Ruc_m} 
\int_{v_{uc}^{(m)}}d \bfrho (\cdots). 
\label{eq:int-over-space-to-sum+int-over-unit-cell}
\end{align}
This should be understood in the following way: Every space point $\rr$ can be reached by first a Bravais lattice vector $\Ruc_m\equiv(\mathcal{X}_m,\mathcal{Y}_m,\mathcal{Z}_m)$  and then a vector $\bfrho$ within the $m^{\textrm{th}}$ unit cell, i.e. $\rr=\Ruc_m+\bfrho$. The superscript $(m)$ on the unit cell volume $v_{uc}^{(m)}$ indicates that the integral is over the $m$th unit cell. Thus, the matrix element is
\begin{align}
&\langle\varphi_{\kk \upsilon\tau}|H_{\textrm{HF},i}|\varphi_{\kk' \upsilon'\tau'}\rangle
\\
&=
\sum_{n}
\sum_{\Ruc_m}\ \int_{v_{uc}^{(m)}}\!d \bfrho\
\varphi_{\kk \upsilon\tau}^\ast(\Ruc_m+\bfrho)h^n_{i}\varphi_{\kk' \upsilon'\tau'}(\Ruc_m+\bfrho).
\nonumber
\end{align}
Here one sum is over \emph{all} unit cells $\Ruc_m$, whereas the other sum is \emph{only} over those atoms at position $\RR_n$ with a non-zero nuclear spin.\cite{footnote-Bravais-lattice-vector-vs-vector-to-atom} To proceed, we take $\upsilon=\upsilon'=H$ as an illustrative example and obtain
\begin{align}
\langle\varphi_{\kk H\tau}|&H_{\textrm{HF},i}|\varphi_{\kk' H\tau'}\rangle
\simeq
\frac{v_a}{L_xL_y}
\sum_{n}\!
\sum_{\Ruc_m}\! 
e^{i(\kk'-\kk)\cdot\Ruc_{m\bot}^{}} 
\\
&\times
|f^{}_H(\mathcal{Z}_m)|^2 
\int_{v_{uc}^{(m)}}\!\!\!\!d \bfrho\
u^\ast_{\Gamma_8,\tau3/2}(\bfrho)
h^n_{i}
u_{\Gamma_8,\tau'3/2}(\bfrho),\nonumber
\end{align}
where we have used the slow variation of the envelope functions on the atomic scale, $f_H(\rho_z+\mathcal{Z}_m)\simeq f_H(\mathcal{Z}_m)$, and the lattice periodicity of the lattice periodic functions, e.g.~$u_{\Gamma_8,\tau\frac{3}{2}}(\Ruc_m+\bfrho)=u_{\Gamma_8,\tau\frac{3}{2}}(\bfrho)$ for all $\Ruc_m$. Here $\Ruc_{m\bot}^{}\equiv(\mathcal{X}_m,\mathcal{Y}_m)$, and the integral over $\bfrho$ is over the $m^{\textrm{th}}$ unit cell, whereas $h^n_i$ is for the $n^{\textrm{th}}$ nuclei.  For a specific nuclear spin $n$, we now include \emph{only} the integral over that particular unit cell containing the $n^{\textrm{th}}$ nuclear spin, since the HF interactions are local in space. In other words, if the nuclei spin $n$ is not inside the integration volume of the unit cell $m$, then the contribution is neglected,\cite{footnote-nabo-spins-not-included} i.e.
\begin{align}
\langle\varphi_{\kk H\tau}&|H_{\textrm{HF},i}|\varphi_{\kk' H\tau'}\rangle
=
\frac{v_a}{L_xL_y}
\sum_{n}
e^{i(\kk'-\kk)\cdot\Ruc_{n\bot}^{}} 
\label{eq:matrix-element-HH-generalt-2D-model}\\
& 
\times
|f^{}_H(\mathcal{Z}_n)|^2
\int_{v_{uc}}\!\! d \bfrho\
u^\ast_{\Gamma_8,\tau3/2}(\bfrho)
h^n_{i}
u^{}_{\Gamma_8,\tau'3/2}(\bfrho),
\nonumber
\end{align}
where the unit cell integral now is independent of the unit cell position $\Ruc_n$. The sum is \emph{only} over the lattice nuclei at $\RR_n$ with a non-zero nuclear spin. Therefore, the system does not have discrete translational symmetry, so the sum cannot simply be made into an integral. Hence, the matrix elements are not diagonal in $\kk$ due to the nuclear spins at random lattice points.

In order to proceed, we need to evaluate the integral of the lattice periodic function over the unit cell in Eq.(\ref{eq:matrix-element-HH-generalt-2D-model}). To this end, the symmetry of the lattice periodic functions are important: The contact interaction $H_{\textrm{HF},1}$ is zero for $P$-like states, since they vanish on the atomic center, while matrix elements of $H_{\textrm{HF},i}$ for $i=2,3$ vanish for $S$-like states due to their spherical symmetry. Here, we approximate the lattice periodic functions by a Linear Combination of Atomic Orbitals (LCAO) as\cite{Gueron-PR-1964,Fischer-Coish-Bulaev-Loss-PRB-2008}  
\begin{subequations}
\label{eq:u6-og-u8-in-the-LCAO-approach}
\begin{align}
u_{\Gamma_6,m_j}^{}(\rr)=&
N_{\Gamma_6,m_j}^{}
\\
&\!\!\times\!\!\Big[\alpha^{}_{\Te}\Psi^{\Te}_{\Gamma_6,m_j}\Big(\rr{+}\frac{\bfd}{2}\Big)
-\alpha^{}_{\Hg}\Psi_{\Gamma_6,m_j}^{\Hg}\Big(\rr{-}\frac{\bfd}{2}\Big)\Big],
\nonumber\\
u_{\Gamma_8,m_j}^{}(\rr)=&
N_{\Gamma_8,m_j}^{}
\\
&\!\!\times\!\!\Big[\alpha^{}_{\Te}\Psi_{\Gamma_8,m_j}^{\Te}\Big(\rr{+}\frac{\bfd}{2}\Big)
+\alpha^{}_{\Hg}\Psi_{\Gamma_8,m_j}^{\Hg}\Big(\rr{-}\frac{\bfd}{2}\Big)\Big],
\nonumber
\end{align}
\end{subequations}
where $\Psi_{\Gamma_i,m_j}^{\Te}$ and $\Psi_{\Gamma_i,m_j}^{\Hg}$ are atomic-like wave functions centered on the Te and Hg atom, respectively, and $\rr$ is only within a single two-atomic primitive unit cell of HgTe centered at $\rr=0$. The atomic wave functions inherit the symmetry of the band\cite{Gueron-PR-1964,Fischer-Coish-Bulaev-Loss-PRB-2008} as indicated by the index $\Gamma_i,m_j$. The atoms are connected by the  vector $\bfd$, and the constants $N_{\Gamma_i,m_j}$ are determined by the lattice periodic function normalization $\int_{v_{uc}}d \rr |u_{\Gamma_i,m_j}(\rr)|^2=2$, see e.g.~Eq.(\ref{eq:lattice-perio-func-normalization}). The electron sharing within the unit cell is described by $\alpha^{}_{\Te(\Hg)}$, which fulfill $|\alpha^{}_{\Te}|^2+|\alpha^{}_{\Hg}|^2=1$.\cite{footnote-bonding-antibonding} 

The LCAO approach (\ref{eq:u6-og-u8-in-the-LCAO-approach}) now facilitates evaluation of the unit cell integral in the matrix elements $\langle\varphi_{\kk \upsilon\tau}|H_{\textrm{HF},i}|\varphi_{\kk' \upsilon'\tau'}\rangle$. Consider e.g. the unit cell integral in Eq.(\ref{eq:matrix-element-HH-generalt-2D-model}) for a non-zero spin on the $n^{\textrm{th}}$ Hg nucleus located on $\bfrho=\bfd/2$, i.e.
\begin{align}
&\int_{v_{uc}}\!\! d \bfrho\
u^\ast_{\Gamma_8,\tau3/2}(\bfrho)
h^n_{i}
u_{\Gamma_8,\tau'3/2}(\bfrho)
\label{eq:int-over-u-to-int-over-orbitals}
\\
&\simeq
N^\ast_{\Gamma_8,\tau3/2}
N_{\Gamma_8,\tau'3/2}
|\alpha^{}_{\Hg}|^2
\nonumber\\
&\times
\int_{v_{uc}}\!\!d \bfrho 
[\Psi_{\Gamma_8,\tau3/2}^{\Hg}(\bfrho-\bfd/2)]^\ast
h^n_{i}
\Psi_{\Gamma_8,\tau'3/2}^{\Hg}(\bfrho-\bfd/2), 
\nonumber
\end{align}
where only the important contribution of the atomic wave functions centered on the Hg atom is included. In other words, integrals involving atomic wave functions centered on different atoms are neglected. Fischer \emph{et al.}\cite{Fischer-Coish-Bulaev-Loss-PRB-2008} estimated that these non-local contributions are two to three orders of magnitude smaller for GaAs -- even for the long-ranged potentials in $h^n_{2,3}$ in Eqs.(\ref{eq:HF2-of-nuclei-nr-n},\ref{eq:HF3-of-nuclei-nr-n}).

Thus, we have now outlined how to find the matrix elements $\langle\varphi_{\kk H\tau}|H_{\textrm{HF},i}|\varphi_{\kk' H\tau'}\rangle$ for all three kinds of HF interactions (\ref{eq:HF123-of-nuclei-nr-n}). The matrix elements of the types  $\langle\varphi_{\kk E\tau}|H_{\textrm{HF},i}|\varphi_{\kk' E\tau'}\rangle$ and $\langle\varphi_{\kk E\tau}|H_{\textrm{HF},i}|\varphi_{\kk' H\tau'}\rangle$ follow the same lines as above. The essential ingredients are the locality of the HF interactions, the periodicity of $u_{\Gamma_i,m_j}(\rr)$  and the slowly varying envelope functions. Next, we find the three HF interactions (\ref{eq:HF123-of-nuclei-nr-n}) within the BHZ model.

\subsection{The contact HF interaction for $S$-like states}

Now we find the contact HF interaction $H_{\textrm{HF},1}$ Eq.(\ref{eq:HF1-of-nuclei-nr-n}) within the BHZ basis (\ref{eq:basis-for-BHZ-2D}). We begin by noting that $\langle\varphi_{\kk H\tau}|H_{\textrm{HF},1}|\varphi_{\kk' H\tau'}\rangle=0$ and $\langle\varphi_{\kk E\tau}|H_{\textrm{HF},1}|\varphi_{\kk' H\tau'}\rangle=0$, since the contact interaction is only non-zero at the atomic center ($h^n_1\propto\delta(\rr-\RR_n)$), where the $\Gamma_8$ $P$-like atomic orbitals vanish. Hence, only the $\Gamma_6$ $S$-like part of the $\varphi_{\kk' E\tau'}(\rr)$ states leads to non-zero matrix elements of $H_{\textrm{HF},1}$. Using the approach in Sec.~\ref{subsec:general-way-to-find-the-matrix-element} to find the matrix elements, we get
\begin{align}
\langle\varphi_{\kk E\tau}|&H_{\textrm{HF},1}|\varphi_{\kk' E\tau'}\rangle
=
\frac{v_a}{L_xL_y}
\sum_{n}
e^{i(\kk'-\kk)\cdot\Ruc_{n\bot}^{}}
\\
&\times 
|f^{}_{E\Gamma_6}(\mathcal{Z}_n)|^2 
\int_{v_{uc}}\!\!d \bfrho\
u^\ast_{\Gamma_6,\tau1/2}(\bfrho)
h^n_{1}
u^{}_{\Gamma_6,\tau'1/2}(\bfrho)
\nonumber
\end{align}
for $\tau,\tau'=\pm$. The $\Gamma_6$ states $u_{\Gamma_6,\pm1/2}(\rr)$ simply factorize into a spin and an orbital part as $u_{\Gamma_6,+(-)1/2}(\rr)=u_{\Gamma_6}(\rr)|\op(\ned)\rangle$, see e.g.~Eq.(\ref{eq:Gamma-states-S}). Using this and the explicit form of the contact interaction $h^n_{1}$ in Eq.(\ref{eq:HF1-of-nuclei-nr-n}), we readily obtain 
\begin{align}
\langle\varphi_{\kk E\tau}|H_{\textrm{HF},1}&|\varphi_{\kk' E\tau'}\rangle
=
\frac{1}{L_xL_y}
\sum_{n}
e^{i(\kk'-\kk)\cdot\Ruc_{n\bot}^{}} 
\label{eq:matrix-element-HF1-final}\\
&\hspace{-4mm}\times
A^{}_{S,j_n}(\mathcal{Z}_n)
\frac{1}{\h}
\left[
\tau\frac{1}{2} I_{z,n}\delta_{\tau,\tau'}
+\frac{1}{2} I_{\tau',n}\delta_{\tau,-\tau'}
\right],
\nonumber
\end{align}
where $I_{\pm,n}\equiv I_{x,n}\pm iI_{y,n}$ are the raising and lowering nuclear spin operators. In analogue to the case of a quantum dot,\cite{Fischer-Coish-Bulaev-Loss-PRB-2008,Coish-Baugh-review-2009} we here introduce the \emph{position dependent contact HF coupling} as\cite{footnote-unit-and-k-dependence-in-HF-constant} 
\begin{align}
A^{}_{S,j_n}(\mathcal{Z}_n)\equiv
v_a
|f^{}_{E\Gamma_6}(\mathcal{Z}_n)|^2 
A^{\textrm{Atomic}}_{S,j_n},
\label{eq:def-HF-position-dependent-contact-coupling}
\end{align}
which includes the \emph{atomic contact HF coupling}
\begin{align}
A^{\textrm{Atomic}}_{S,j_n}\equiv
\frac{2\mu_0}{3} 
g_e\mu_Bg_{j_n}\mu_N
|u^{}_{\Gamma_6}(\RR_n)|^2
\label{eq:atomic-HF-couplings-def}
\end{align}
for the nuclear spin at site $n$ of isotope $j$. Here $A^{}_{S,j_n}(\mathcal{Z}_n)$ depends on the real space position of the nuclear spin. In contrast, $A^{\textrm{Atomic}}_{S,j_n}$ does \emph{not} depend on the nuclear position, since it can be given in terms of the atomic orbital $\Psi^{j_n}_{\Gamma_6}$ by using Eq.(\ref{eq:u6-og-u8-in-the-LCAO-approach}) as $A^{\textrm{Atomic}}_{S,j_n}\propto |u^{}_{\Gamma_6}(\RR_n)|^2 \simeq |N_{\Gamma_6,1/2}^{}|^2 |\alpha^{}_{j_n}|^2|\Psi^{j_n}_{\Gamma_6}(0)|^2$, i.e. $A^{\textrm{Atomic}}_{S,j_n}$ only depends on the nuclear isotope type $j_n$ at site $n$. Moreover, at the present level of approximation, we can freely replace the Bravais lattice vector $\Ruc_{n}$ by the actual position of a nuclear spin within the $n^{\textrm{th}}$ unit cell in the envelope functions in Eq.(\ref{eq:matrix-element-HF1-final}) due to their slow variation. Finally, we arrive at the HF contact interaction in the BHZ basis as 
\begin{align}
\mathcal{H}_{\textrm{HF},1}= 
\sum_{n}
\sum_{\kk,\kk'} 
\frac{e^{i(\kk'-\kk)\cdot\Ruc_{n\bot}^{}}}{L_xL_y}\
 \mathbf{c}_{\kk}^\dag \tilde{H}_{\textrm{HF},1} \mathbf{c}_{\kk'}^{}
\end{align}
where $\mathbf{c}_{\kk}^\dag=(c_{\kk,E+}^\dag,c_{\kk,H+}^\dag,c_{\kk,E-}^\dag,c_{\kk,H-}^\dag)$ and\cite{footnote-prefactor-HF1} 
\begin{align}
\tilde{H}_{\textrm{HF},1}=
\frac{1}{2\h}
A^{}_{S,j_n}(\mathcal{Z}_n)
\left(
\begin{array}{cccc}
 I_{z,n} & 0 & I_{-,n} & 0 \\
 0 & 0 & 0 & 0 \\
 I_{+,n} & 0 & -I_{z,n} & 0 \\
 0 & 0 & 0 & 0
\end{array}
\right).
\label{eq:H-HF-contact-BHZ-model}
\end{align}
The sum is only over non-zero nuclear spins. Therefore, it is now clear that the contact HF interaction contains elements $\propto I_{\pm,n}$, which connect the time-reversed blocks in the BHZ hamiltonian (\ref{eq:H-BHZ-model-matrix1}). Moreover, as illustrated in Fig.\ref{fig:Spin-flips-between-BHZ-states}, only the $|E\pm\rangle$ states are connected by $H_{\textrm{HF},1}$, since only these states contain a $S$-like symmetry part. In in table \ref{Table:atomic-S-HF-couplings}, estimates of the atomic contact HF couplings $A^{\textrm{Atomic}}_{S,j_n}$ are given for the stable isotopes of HgTe with non-zero nuclear spin (see Appendix \ref{App:estimation-of-HF-constants} for details).

\begin{table}
  \begin{tabular}{ l | c | c | c | c | }
     & $^{199}$Hg & $^{201}$Hg & $^{123}$Te & $^{125}$Te \\ 
    \hline
    $A^{\textrm{Atomic}}_{S,j_n}$ [$\mu$eV] & 4.1 & -1.5 & -49 & -59 \\
    \hline
     $A^{\textrm{Atomic}}_{P,j_n}$ [$\mu$eV] & 0.6 & -0.2 & -6.0 & -7.2 \\ 
    \hline
  \end{tabular}
\caption{Estimates of the atomic contact HF couplings $A^{\textrm{Atomic}}_{S,j_n}$ Eq.(\ref{eq:atomic-HF-couplings-def}) and the atomic $P$-like HF couplings $A^{\textrm{Atomic}}_{P,j_n}$ Eq.(\ref{eq:atomic-P-HF-coupling-def}) in HgTe for the naturally present isotopes with non-zero spin, see (\ref{eq:isotopes-of-HgTe}). The HF couplings for $S$-like states are seen to be about one order of magnitude larger than for $P$-like states. The sign of the HF couplings stems from the sign of the nuclear $g$-factors. See Appendix \ref{App:estimation-of-HF-constants} for details of these estimates.}
\label{Table:atomic-S-HF-couplings}
\end{table}

\subsection{The HF interactions for $P$-like states}\label{subsec:HF-for-P-states}

Next, we find the HF interactions within the BHZ basis (\ref{eq:basis-for-BHZ-2D}) for $H_{\textrm{HF},2}$ and $H_{\textrm{HF},3}$ Eqs.(\ref{eq:HF2-of-nuclei-nr-n},\ref{eq:HF3-of-nuclei-nr-n}), which are relevant for the $P$-like states. 

To begin with, we argue that the $\Gamma_6$ $S$-like states -- part of the $E\pm$ states -- do not contribute to the matrix elements $\langle\varphi_{\kk E\tau}|H_{\textrm{HF},i}|\varphi_{\kk' E\tau'}\rangle$ and $\langle\varphi_{\kk E\tau}|H_{\textrm{HF},i}|\varphi_{\kk' H\tau'}\rangle$ for $i=2,3$. (In contrast, the $\Gamma_8$ $P$-like part of $E\pm$ do contribute to these elements as will be shown below.) To understand this, the HF matrix elements are written in terms of the unit cell integrals over the atomic-like wave functions as outlined in Sec.~\ref{subsec:general-way-to-find-the-matrix-element}. Firstly, for the dipole-dipole like HF interaction (\ref{eq:HF2-of-nuclei-nr-n}), we have 
\begin{align}
\int_{v_{uc}} \!\!\! d \bfrho
\big[\Psi^{\Hg/\Te}_{\Gamma_6,m_j}(\bfrho\mp\bfd/2)\big]^\ast
h^n_{2}
\Psi^{\Hg/\Te}_{\Gamma_6,m'_j}(\bfrho\mp\bfd/2)=0 
\end{align}
due to the rotational symmetry of the $S$-like orbitals around the atomic core.\cite{footnote-r_c-cut-off} Secondly, we have 
\begin{align}
\int_{v_{uc}} \!\!\! d \bfrho
[\Psi^{\Hg/\Te}_{\Gamma_6,m_j}(\bfrho\mp\bfd/2)]^\ast
h^n_{2}
\Psi^{\Hg/\Te}_{\Gamma_8,m'_j}(\bfrho\mp\bfd/2)=0
\end{align}
due to opposite parities of the $S$- and $P$-like orbitals.\cite{footnote-atomic-parity} The same matrix elements containing $h^n_{3}$ instead of $h^n_{2}$ are also zero, because the $S$-like states have zero orbital momentum, i.e.~$\mathbf{L}_n\Psi^{\Hg/\Te}_{\Gamma_6,m_j}(\bfrho)=0$. 

Therefore, only $P$-like states contribute, so we are now left with (see Sec.~\ref{subsec:general-way-to-find-the-matrix-element})
\begin{subequations}
\label{eq:P-matrix-elem-mellem1}
\begin{align}
\langle\varphi_{\kk E\tau}|H_{\textrm{HF},i}|\varphi_{\kk' E\tau'}\rangle
&=
\frac{v_a}{L_xL_y}
\sum_{n}
e^{i(\kk'-\kk)\cdot\Ruc_{n\bot}^{}} 
\\
\times
|f_{E\Gamma_8}(\mathcal{Z}_n)|^2 
&\int_{v_{uc}}\!\! d \bfrho
u^\ast_{\Gamma_8,\tau1/2}(\bfrho)
h^n_{i}
u_{\Gamma_8,\tau'1/2}(\bfrho),
\nonumber\\
\langle\varphi_{\kk H\tau}|H_{\textrm{HF},i}|\varphi_{\kk' H\tau'}\rangle
&=
\frac{v_a}{L_xL_y}
\sum_{n}
e^{i(\kk'-\kk)\cdot\Ruc_{n\bot}^{}}
\\
\times 
|f_H(\mathcal{Z}_n)|^2 
&\int_{v_{uc}}\!\! d \bfrho\
u^\ast_{\Gamma_8,\tau3/2}(\bfrho)
h^n_{i}
u_{\Gamma_8,\tau'3/2}(\bfrho),
\nonumber\\
\langle\varphi_{\kk E\tau}|H_{\textrm{HF},i}|\varphi_{\kk' H\tau'}\rangle
&=
\frac{v_a}{L_xL_y}
\sum_{n}
e^{i(\kk'-\kk)\cdot\Ruc_{n\bot}^{}}
\\
\times 
f^{\ast}_{E\Gamma_8}(\mathcal{Z}_n)f_H(\mathcal{Z}_n)
&\int_{v_{uc}}\!\!d \bfrho\
u^\ast_{\Gamma_8,\tau1/2}(\bfrho)
h^n_{i}
u_{\Gamma_8,\tau'3/2}(\bfrho)
\nonumber
\end{align}
\end{subequations}
and $\langle\varphi_{\kk H\tau}|H_{\textrm{HF},i}|\varphi_{\kk' E\tau'}\rangle=\langle\varphi_{\kk' E\tau'}|H_{\textrm{HF},i}|\varphi_{\kk H\tau}\rangle^\ast$, where $i=2,3$ and $\tau,\tau'=\pm$. Using the LCAO approach  Eq.(\ref{eq:u6-og-u8-in-the-LCAO-approach}), the unit cell integrals over the lattice periodic functions now become integrals over the atomic-like wave functions as in Eq.(\ref{eq:int-over-u-to-int-over-orbitals}). We write the atomic wave functions as a product of a radial part $R^{\Hg/\Te}(r)$ and an angular part $\mathbb{Y}_{\Gamma_8,m_j}(\theta,\phi)$, i.e. $\Psi_{\Gamma_8,m_j}^{\Hg/\Te}(\rr)=R^{\Hg/\Te}(r)\mathbb{Y}_{\Gamma_8,m_j}(\theta,\phi)$, using spherical coordinates $(r,\theta,\phi)$ with the nucleus in the center. Since the integrals are over the two-atomic unit cell volume, they do not \emph{a priori} factorize into a product of radial and angular integrals. However, due to the $1/r^3$ dependence of $h^n_{i}$ ($i=2,3$), the important part of the unit cell integrals are numerically within one or two Bohr radii $a_0$ from the atomic core, which is certainly within the unit cell volume. Therefore, it is a good approximation to write the unit cell integrals (e.g.~Eq.(\ref{eq:int-over-u-to-int-over-orbitals})) as
\begin{align}
\int_{v_{uc}}\!\!\!\!& d \bfrho 
[\Psi_{\Gamma_8,m_j^{}}^{\Hg}(\bfrho-\bfd/2)]^\ast
h^n_{i}
\Psi_{\Gamma_8,m'_j}^{\Hg}(\bfrho-\bfd/2) 
\label{eq:speherical-approx-for-unit-cell-int-over-h2-ogh3}\\
&\simeq\int_0^{r_{\textrm{max}}}\!\!\!\!\!\!\! d r r^2
\! \int_0^{2\pi} \!\!\!\!\!\!\!d \phi
\! \int_0^{\pi}\!\!\!\!\!d \theta \sin(\theta)  
[\Psi_{\Gamma_8,m_j^{}}^{\Hg}(\rr)]^\ast
h^n_{2}
\Psi_{\Gamma_8,m'_j}^{\Hg}(\rr),
\nonumber
\end{align}
where the specific choice of $r_{\textrm{max}}\gtrsim a_0$ is not important for the numerical value of the integral.\cite{footnote-chosing-r_max} Therefore, we are now left with an essentially atomic physics problem, where the integral separates into a product of a radial and an angular part. The radial part is 
\begin{align}
\left\langle\frac{1}{r^3}\right\rangle^{\kappa}_{r}
\equiv
\int_0^{r_{\textrm{max}}}\!\!\! d r  r^2
|R^{\kappa}(r)|^2
\frac{1}{r^3(1+\frac{r_c}{r})}, 
\label{eq:def-<1/r^3>}
\end{align}
which is the same for all the matrix elements of $h^n_{2}$ and $h^n_{3}$ and only depends on the type of atom $\kappa=$Hg, Te. Due to the smallness of the nuclear length scale $r_c$, it is not significant for the magnitude of $\langle1/r^3\rangle^{\kappa}_{r}$.\cite{footnote-r_c-cut-off} Using the radial integral (\ref{eq:def-<1/r^3>}), we introduce the \emph{atomic} $P$\emph{-like HF coupling} for isotope $j$ (at site $n$) as
\begin{align}
A^{\textrm{Atomic}}_{P,j_n}\equiv
\frac{\mu_0}{4\pi}
g^{}_e\mu^{}_Bg^{}_{j_n}\mu^{}_N
(N_{\Gamma_8})^2|\alpha_{j_n}|^2
\left\langle\frac{1}{r^3}\right\rangle^{j_n}_{r},
\label{eq:atomic-P-HF-coupling-def}
\end{align}
which are estimated to be about one order of magnitude smaller than the atomic contact HF couplings $A^{\textrm{Atomic}}_{S,j_n}$ (\ref{eq:atomic-HF-couplings-def}), see table \ref{Table:atomic-S-HF-couplings}. Here it makes sense to have a common atomic HF coupling for the dipole-dipole like coupling $h_{2}^n$ and the orbital to nuclear-spin coupling $h_3^n$, since the normalization constants for the LCAO lattice functions (\ref{eq:u6-og-u8-in-the-LCAO-approach}) are numerically approximately equal, $N_{\Gamma_8,3/2}\simeq N_{\Gamma_8,1/2}\equiv N_{\Gamma_8}$, as discussed in Appendix \ref{App:estimation-of-HF-constants}.\cite{footnote-different-P-HF-couplings-possible} Moreover, we also use that $N_{\Gamma_i,m_j}$ are independent of the sign of $m_j$, see Eq.(\ref{eq:normalization-consts-uaf-af-fortegn-af-mj}). Calculating the angular integrals as discussed in Appendix \ref{App:atomic-integrals}, the matrix elements (\ref{eq:P-matrix-elem-mellem1}) for the dipole-dipole like HF interaction $H_{\textrm{HF},2}$  become
\begin{subequations}
\label{eq:matrix-elem-HF2-final}
\begin{align}
\langle\varphi_{\kk E\tau}|H_{\textrm{HF},2}|\varphi_{\kk' E\tau'}\rangle
=&
\sum_{n}
\frac{1}{L_xL_y} e^{i(\kk'-\kk)\cdot\Ruc_{n\bot}^{}}
\label{eq:matrix-elem-HF2-EE}\\
&\hspace{-2cm}
\times
A^{EE}_{P,j_n}
\frac{1}{\h}
\Big[
-\tau\delta^{}_{\tau',\tau}
\frac{1}{15} I_{z,n}
-
\delta^{}_{\tau',-\tau}
\frac{2}{15} I_{-\tau,n}
\Big],
\nonumber\\
\langle\varphi_{\kk H\tau}|H_{\textrm{HF},2}|\varphi_{\kk' H\tau'}\rangle
=&
\sum_{n}
\frac{1}{L_xL_y}e^{i(\kk'-\kk)\cdot\Ruc_{n\bot}^{}}
\label{eq:matrix-elem-HF2-HH}\\
&\hspace{6mm}\times 
A^{HH}_{P,j_n}
\delta_{\tau,\tau'}
\frac{1}{\h}
(-\tau)\frac{1}{5} I_{z,n},
\nonumber\\
\langle\varphi_{\kk E\tau}|H_{\textrm{HF},2}|\varphi_{\kk' H\tau'}\rangle
=&\!
\sum_{n}\!
\frac{e^{i(\kk'-\kk)\cdot\Ruc_{n\bot}^{}}}{L_xL_y}
\frac{{-}A^{EH}_{P,j_n}\!\delta^{}_{\tau',\tau}I_{\tau,n}}{5\sqrt{3}\h}
\end{align}
\end{subequations}
where we introduce the \emph{position dependent} $P$-like HF couplings as  
\begin{subequations}
\label{eq:position-dependent-P-HF-coulings-def}
\begin{align}
A^{HH}_{P,j_n}
&\equiv
v_a |f^{}_{H}(\mathcal{Z}_n)|^2 
A^{\textrm{Atomic}}_{P,j_n},\\
A^{EE}_{P,j_n}
&\equiv
v_a |f^{}_{E\Gamma_8}(\mathcal{Z}_n)|^2 
A^{\textrm{Atomic}}_{P,j_n},\\
A^{EH}_{P,j_n}
&\equiv
v_a
f^{\ast}_{E\Gamma_8}(\mathcal{Z}_n)
f^{}_{H}(\mathcal{Z}_n)
A^{\textrm{Atomic}}_{P,j_n},
\end{align}
\end{subequations} 
and  $A^{HE}_{P,j_n}=[A^{EH}_{P,j_n}]^\ast$. In comparison, for the contact HF interaction only a single position dependent HF coupling was introduced in Eq.(\ref{eq:def-HF-position-dependent-contact-coupling}). Here the explicit dependence on the position $\mathcal{Z}_n$ of the nuclear spin has been suppressed in the notation for simplicity, i.e. $A^{\upsilon\upsilon'}_{P,j_n}(\mathcal{Z}_n)=A^{\upsilon\upsilon'}_{P,j_n}$. Similarly, the matrix elements for the HF interaction $H_{\textrm{HF},3}$ between the electronic orbital momentum and the nuclear spins become
\begin{subequations}
\label{eq:matrix-elem-HF3-final}
\begin{align}
\langle\varphi_{\kk E\tau}|H_{\textrm{HF},3}|\varphi_{\kk' E\tau'}\rangle
=&
\sum_{n}
\frac{1}{L_xL_y}e^{i(\kk'-\kk)\cdot\Ruc_{n\bot}^{}} 
\label{eq:matrix-elem-HF3-EE}\\
&\hspace{-1.cm}\times
A_{P,j_n}^{EE}
\frac{1}{\h}
\Big[
\delta_{\tau,\tau'}
\frac{1}{3}\tau I_{z,n}
+
\delta_{-\tau,\tau'}
\frac{2}{3}I_{-\tau,n}
\Big],
\nonumber\\
\langle\varphi_{\kk H\tau}|H_{\textrm{HF},3}|\varphi_{\kk' H\tau'}\rangle
=&
\sum_{n}
\frac{e^{i(\kk'-\kk)\cdot\Ruc_{n\bot}^{}}}{L_xL_y}
\frac{A_{P,j_n}^{HH}\delta_{\tau,\tau'}\tau I_{z,n}}{\h},
\label{eq:matrix-elem-HF3-HH}
\\
\langle\varphi_{\kk E\tau}|H_{\textrm{HF},3}|\varphi_{\kk' H\tau'}\rangle
=&
\sum_{n}
\frac{e^{i(\kk'-\kk)\cdot\Ruc_{n\bot}^{}} }{L_xL_y}
\frac{A_{P,j_n}^{EH}\delta_{\tau,\tau'}I_{\tau,n}}{\sqrt{3}\h}.
\end{align}
\end{subequations}
It is noteworthy that the heavy hole like states $H\pm$ only couple diagonally ($\tau=\tau'$) or Ising-like in (\ref{eq:matrix-elem-HF2-HH},\ref{eq:matrix-elem-HF3-HH}) in agreement with Ref.[\onlinecite{Fischer-Coish-Bulaev-Loss-PRB-2008}]. Physically, this is because the $H\pm$ states have a difference of total angular momentum projection larger than one, $|m_j-m_{j'}|>1$. Moreover, the coupling between the states $E\pm$ in Eqs.(\ref{eq:matrix-elem-HF2-EE},\ref{eq:matrix-elem-HF3-EE}) is essentially like the coupling between the light hole states $|\Gamma_8,\pm1/2\rangle$, since the $S$-like states do not contribute to the matrix elements of $H_{\textrm{HF},2}$ and $H_{\textrm{HF},3}$. For these matrix elements between the $E\pm$ states, the off-diagonal elements ($\tau=-\tau'$) are a factor of 2 larger than the diagonal elements ($\tau=\tau'$) in accordance with Ref.[\onlinecite{Testelin-Bernardot-Eble-Chamarro-PRB-2009}].  

Using the matrix elements in Eqs.(\ref{eq:matrix-elem-HF2-final},\ref{eq:matrix-elem-HF3-final}), we now finally arrive at the HF interactions relevant for the $P$-like states in the basis (\ref{eq:basis-for-BHZ-2D}) as 
\begin{align}
\mathcal{H}_{\textrm{HF},i}= 
\sum_{n}
\sum_{\kk,\kk'} 
\frac{e^{i(\kk'-\kk)\cdot\Ruc_{n\bot}^{}}}{L_xL_y}
\mathbf{c}_{\kk}^\dag \tilde{H}_{\textrm{HF},i} \mathbf{c}_{\kk'}^{}
\end{align}
for $i=2,3$, where
\begin{subequations}
\label{eq:H-HF-both-P-like-BHZ-model}
\begin{widetext}
\begin{align}
\tilde{H}_{\textrm{HF},2}&=
\frac{1}{5\h}
\left(
\begin{array}{cccc}
 -\frac{1}{3}A^{EE}_{P,j_n} I_{z,n}  
 & -\frac{1}{\sqrt{3}}A^{EH}_{P,j_n}I_{+,n} 
 & -\frac{2}{3}A^{EE}_{P,j_n} I_{-,n} & 0 \\
 -\frac{1}{\sqrt{3}}A^{HE}_{P,j_n}I_{-,n} 
 & -A^{HH}_{P,j_n} I_{z,n} & 0 & 0 \\
 -\frac{2}{3}A^{EE}_{P,j_n} I_{+,n} 
 & 0 & \frac{1}{3}A^{EE}_{P,j_n} I_{z,n} 
 & -\frac{1}{\sqrt{3}}A^{EH}_{P,j_n}I_{-,n} \\
 0 & 0 & -\frac{1}{\sqrt{3}}A^{HE}_{P,j_n}I_{+,n} 
 & A^{HH}_{P,j_n}I_{z,n}
\end{array}
\right),
\label{eq:H-HF-dipole-BHZ-model}
\end{align}
\end{widetext}
and
\begin{align}
\tilde{H}_{\textrm{HF},3}=
-5\tilde{H}_{\textrm{HF},2}
\label{eq:H-HF-3-BHZ-model}
\end{align}
\end{subequations}
such that the total HF interaction for the $P$-like states becomes
\begin{align}
\tilde{H}_{\textrm{HF},P}
=
\tilde{H}_{\textrm{HF},2}+\tilde{H}_{\textrm{HF},3}
=-4\tilde{H}_{\textrm{HF},2}.
\label{eq:H-HF-P-BHZ-model}
\end{align}
Just as the contact HF interaction (\ref{eq:H-HF-contact-BHZ-model}), the $P$-like HF interaction connects the time-reversed blocks by connecting the $E\pm$ states. Moreover, the $P$-like HF interaction connects the states \emph{within} the time-reversed blocks (e.g.~$E+$ and $H+$) in contrast to the contact HF interaction, see Fig.~\ref{fig:Spin-flips-between-BHZ-states}. 

Interestingly, the sign of the dipole-dipole like HF interaction (\ref{eq:H-HF-dipole-BHZ-model}) is opposite to the contact HF interaction (\ref{eq:H-HF-contact-BHZ-model}) and to the orbital to nuclear-spin coupling (\ref{eq:H-HF-3-BHZ-model}). However, since the elements of $\tilde{H}_{\textrm{HF},3}$ are larger than those of $\tilde{H}_{\textrm{HF},2}$ in absolute value, the total HF interaction for the $P$-like states (\ref{eq:H-HF-P-BHZ-model}) ends up having the same sign as the contact HF interaction.

\section{Hyperfine interactions for a nanostructure in a $\textrm{HgTe}$ quantum well}\label{sec:HF-any-nanostruc}

Now, we show how the HF interactions for \emph{any} nanostructure in a HgTe QW can be derived from our results in Eqs.(\ref{eq:H-HF-contact-BHZ-model},\ref{eq:H-HF-both-P-like-BHZ-model}) for a HgTe QW with periodic boundary conditions. Examples of such structures are quantum dots,\cite{Chang-Lou-PRL2011} mesoscopic rings,\cite{Michetti-Recher-2011} point contacts\cite{Zhang-et-al-PRB-2011} and anti-dots.\cite{Shan-et-al-PRB-2011} 

For a given nanostructure, the envelope wave functions are needed in order to find its HF interactions within the BHZ framework. Utilizing the Peierls substitution $(k_x,k_y)=-i(\p_x,\p_y)$ in the BHZ hamiltonian (\ref{eq:H-BHZ-model-matrix1}), the envelope functions $\Phi^{}_{\eta}(\rr_{\bot})$ can be found by solving $[H_0(-i\p_x,-i\p_y)+V(\rr_{\bot})]\Phi^{}_{\eta}(\rr_{\bot})=E_\eta^{}\Phi^{}_{\eta}(\rr_{\bot})$, where $V(\rr_{\bot})$ is the potential confining the nanostructure,\cite{Rothe-Hankiewicz-Trauzettel-Guigou-PRB-2012} $\rr_{\bot}\equiv (x,y)$ and $\eta$ is a collection of quantum numbers to be specified for a concrete situation.\cite{footnote-confinement-of-nanostructures,Zhou-edge-states-PRL-2008,Chang-Lou-PRL2011,Michetti-Recher-2011,Shan-et-al-PRB-2011} Terms related to bulk inversion asymmetry,\cite{Konig-JPSJ-review-2008,Delplace-Li-Buttiker-PRL-2012} Rashba spin-orbit coupling\cite{Hankiewicz-NJP-2010,Schmidt-Rachel-Oppen-Glazman-PRL-2012} and/or magnetic fields\cite{Scharf-Matos-Abiague-Fabian-PRB-2012,Hankiewicz-PRL-2010} can also be included here. The envelope function is given by\cite{footnote-normalization-general}
\begin{align}
\Phi^{}_{\eta}(\rr_{\bot})=
\left(
\begin{array}{c}
\phi^{}_{\eta,E+}(\rr_{\bot})\\
\phi^{}_{\eta,H+}(\rr_{\bot})\\
\phi^{}_{\eta,E-}(\rr_{\bot})\\
\phi^{}_{\eta,H-}(\rr_{\bot})
\end{array}
\right)
\label{eq:envelope-4-vector-general} 
\end{align}
such that the entire wave function including the lattice periodic functions is
\begin{align}
\psi_\eta(\rr)=
\sqrt{v_a}\!\!\!\!
\sum_{\zeta=E\pm,H\pm}
\!\!\!\!
\phi^{}_{\eta,\zeta}(\rr_{\bot})\langle \rr|\zeta\rangle,
\end{align}
where $\langle \rr|E\pm\rangle=f_{E\Gamma_6}(z)u_{\Gamma_6,\pm\frac{1}{2}}(\rr)+f_{E\Gamma_8}(z)u_{\Gamma_8,\pm\frac{1}{2}}(\rr)$ and $\langle \rr|H\pm\rangle=f_H(z)u_{\Gamma_8,\pm\frac{3}{2}}(\rr)$. For instance, the 2D QW with periodic boundary conditions simply has the envelope functions $e^{i(k_xx+k_yy)}(1,0,0,0)^T/\sqrt{L_xL_y}$ etc.,~see Eq.(\ref{eq:basis-for-BHZ-2D}).

To find the HF interactions (\ref{eq:HF123-of-nuclei-nr-n}) for a given nanostructure with envelope wave function $\Phi^{}_{\eta}(\rr_\bot)$ (\ref{eq:envelope-4-vector-general}), the same idea of separation of length scales as in Sec.~\ref{sec:HF-in-BHZ} is used: The HF interactions act on the atomic length scale such that the slowly varying envelope functions only become multiplicative factors in the HF interactions for the nanostructure. Thus, we find
\begin{align}
\mathcal{H}_{\textrm{HF},i}^{}= 
\sum_{n}
\sum_{\eta,\eta'} 
[\Phi^{}_{\eta}(\Ruc_{n\bot})]^\dag
\tilde{H}_{\textrm{HF},i}^{}
\Phi^{}_{\eta'}(\Ruc_{n\bot})
c_{\eta}^\dag c_{\eta'}^{}, 
\label{eq:HF-for-any-nanostructure}
\end{align}
where  $\tilde{H}_{\textrm{HF},i}$ are the $4\times4$ matrices found in Eqs.(\ref{eq:H-HF-contact-BHZ-model},\ref{eq:H-HF-both-P-like-BHZ-model},\ref{eq:H-HF-P-BHZ-model}) for the contact ($i=1$) and $P$-like ($i=2,3,P$) HF interactions, respectively. The sum is only over the atomic sites $n$ with a non-zero nuclear spin. The Bravais lattice vector $\Ruc_{n\bot}^{}\equiv(\mathcal{X}_n,\mathcal{Y}_n)$ pointing to the unit cell containing the $n^{\textrm{th}}$ nuclear spin can freely be interchanged by the atomic position $\RR_n$ of the nuclear spin due to the slow variation of the envelope functions on the atomic scale.\cite{footnote-Bravais-lattice-vector-vs-vector-to-atom} 

In situations with time-reversal symmetry, the states appear in Kramers pairs of equal energy. The BHZ model in Eq.(\ref{eq:H-BHZ-model}) is constructed such that Kramers pairs appear as equal energy solutions of the upper and lower $2\times2$ blocks in $H_0$. Thus, $\phi^{}_{\eta,E-}=\phi^{}_{\eta,H-}=0$ for one of the two states in a Kramers pair and vice versa, which simplifies the algebraic burden of finding $\Phi^{}_{\eta}$.\cite{Zhou-edge-states-PRL-2008,Chang-Lou-PRL2011,Michetti-Recher-2011,Shan-et-al-PRB-2011} The two states in a Kramers pair are sometimes referred to as spin up and down, since the upper (lower) block only consists of orbital states with positive (negative) total angular momentum projection, see Eq.(\ref{eq:mz-of-BHZ-states}). If other time-reversal invariant interactions such as the Rashba spin-orbit coupling\cite{Hankiewicz-NJP-2010,Schmidt-Rachel-Oppen-Glazman-PRL-2012} or bulk inversion asymmetry terms\cite{Konig-JPSJ-review-2008} are included into the BHZ model for a nanostructure, then the states still appear in Kramers pairs -- even though the Hamiltonian is \emph{not} necessarily in block diagonal form anymore. In this case, the general formula (\ref{eq:HF-for-any-nanostructure}) for the HF interactions remains valid, since only the slowly varying envelope functions are affected. Here the index for the Kramers pair is included in $\eta$.

Thus, we have now provided the general form of the HF interactions in the BHZ model for a given nanostructure in Eq.(\ref{eq:HF-for-any-nanostructure}). Below we illustrate its use by an example.

\section{Hyperfine interactions for a pair of helical edge states}\label{sec:HF-for-HES}

Next, we deal with the HF interactions for a pair of HESs in a HgTe 2D TI QW. 

\subsection{The helical edge states along the $y$-axis}\label{sec:HES}

\begin{figure}
\includegraphics[width=0.42\textwidth,angle=0]{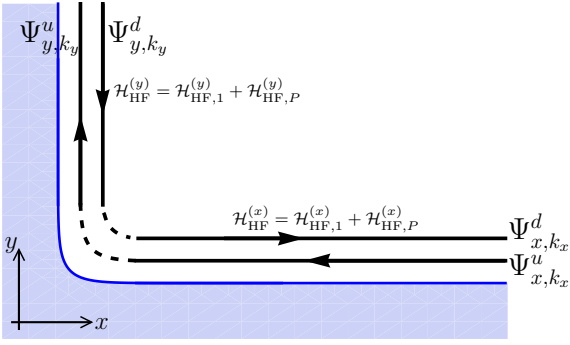} 
\caption{The helical edge states along the $y$-axis Eq.(\ref{eq:HES-along-y-axis}) and along the $x$-axis Eq.(\ref{eq:HES-along-x-axis}) are connected by $k_y\leftrightarrow -k_x$. Moreover, they differ by an imaginary unit $i$ in the $E$ spinor components [compare Eqs.(\ref{eq:4-vector-for-HES-along-y-axis}) and (\ref{eq:4-vector-for-HES-along-x-axis})]. This leads to interesting differences in their HF Hamiltonians (see the main text). For illustrative purposes, the HESs are drawn side by side even though equal energy HESs in fact are \emph{on top} of each other.}
\label{fig:HES-corner}
\end{figure}

To find the HF interactions, we first give the envelope wave functions for a pair of HESs. These can be found by introducing a boundary in the BHZ model and requiring that the envelope functions vanish at the boundary.\cite{Zhou-edge-states-PRL-2008} For a semi-infinite half plane restricted to $x>0$ and periodic boundary condition in the $y$-direction, $k_y$ is still a good quantum number. The HESs envelope functions running along the $y$-direction become\cite{Zhou-edge-states-PRL-2008} (see Fig.\ref{fig:HES-corner})  
\begin{align}
\label{eq:HES-along-y-axis}
\Psi^{\varsigma}_{y,k_y}(x,y)&=
\frac{1}{\sqrt{L_y}}
e^{ik_yy}
\mathfrak{h}_{k_y}^{\varsigma}(x)
\chi_{y}^{\varsigma}
\quad\textrm{for}\quad
\varsigma=u,d
\end{align}
with the energy dispersions $E^u_{k_y}=E_0+\h v_0 k_y$ and $E^d_{k_y}=E_0-\h v_0 k_y$, respectively. The dispersions are exactly linear in the semi-infinite half plane model used here.\cite{Bihlmayer-Edge-states-in-Bi-films-PRB-2011} Both the velocity $v_0=-\sqrt{B^2-D^2}|A|/(\h B)$ and $E_0=-M_0D/B$ are positive for realistic parameters\cite{footnote-parameters-for-BHZ-model} and $L_y$ is the length of the edge. The spinor parts of the HESs are independent of $k_y$ and given by
\begin{align}
\chi_{y}^{u}=
\mathfrak{n}
\left(
\begin{array}{c}
 -i\frac{A}{|A|}\\
 \frac{\sqrt{B^2-D^2}}{B-D}\\
 0\\
 0
\end{array}
\right),
\quad
\chi_{y}^{d}=
\mathfrak{n}
\left(
\begin{array}{c}
 0\\
 0\\
 +i\frac{A}{|A|}\\
 \frac{\sqrt{B^2-D^2}}{B-D}
\end{array}
\right),
\label{eq:4-vector-for-HES-along-y-axis}
\end{align}
with the normalization factor $\mathfrak{n}=\sqrt{(B-D)/2B}$. The transverse part of the HESs are $\mathfrak{h}_{k_y}^{u}(x)=g^{}_{+k_y}(x)$ and $\mathfrak{h}_{k_y}^{d}(x)=g_{-k_y}^{}(x)$, where 
\begin{subequations}
\label{eq:transverse-ES-wave-function-including-lenghts}
\begin{align}
\label{eq:transverse-ES-wave-function}
g^{}_{k_y}(x)=
\sqrt{\frac{2\lambda_1\lambda_2(\lambda_1+\lambda_2)}{(\lambda_1-\lambda_2)^2}}
\left(e^{-\lambda_1 x}-e^{-\lambda_2 x}\right), 
\end{align}
with the $k_y$-dependence inside $\lambda_{1}$ and $\lambda_{2}$ as 
\begin{align}
\lambda_1
&=\frac{1}{\sqrt{B^2-D^2}} \left(\frac{|A|}{2}+\sqrt{W_{k_y}}\right),\\
\lambda_2
&=\frac{1}{\sqrt{B^2-D^2}} \left(\frac{|A|}{2}-\sqrt{W_{k_y}}\right).
\end{align}
Here $\lambda_2^{-1}$ determines the decay length scale of the HES into the bulk and  
\begin{align}
W_{k_y}=&
\left[\frac{A^2}{4}-\frac{M_0}{B}(B^2-D^2)\right]
\nonumber\\
&+\frac{D|A|\sqrt{B^2-D^2}}{B}k_y
+[B^2-D^2]k_y^2. 
\end{align}  
\end{subequations}
The HESs only exist in the topological regime of the BHZ model where $M_0/B>0$. The explicit forms above were derived under the assumption $0\leq M_0/B\leq A^2/(4B^2)$, where $\lambda_{1,2}$ are purely real.\cite{Zhou-edge-states-PRL-2008,footnote-complex-lambdas} This is the relevant regime for the realistic parameters\cite{Qi-Zhang-review-RMP-2010,footnote-parameters-for-BHZ-model} for 2D TI in a HgTe QW of width 61\AA\ or 70\AA. 

Using the time-reversal properties of the basis states of the BHZ model (as discussed in Appendix \ref{App:TR-af-basis-states}), it is seen explicitly that $\Psi^u_{y,k_y}(x,y)$ and $\Psi^d_{y,-k_y}(x,y)$ constitute a Kramers pair, since they are connected by the time-reversal operator $\Theta$ as $\Theta\Psi^u_{y,k_y}(x,y)=-\Psi^d_{y,-k_y}(x,y)$ and $\Theta\Psi^d_{y,k_y}(x,y)=\Psi^u_{y,-k_y}(x,y)$. Often\cite{Qi-Zhang-review-RMP-2010,Konig-JPSJ-review-2008} $\Psi^u_{y,k_y}$ ($\Psi^d_{y,k_y}$) is referred to as the spin-up (spin-down) edge state, since it only consists of states with positive (negative) total angular momentum projection, see Eq.(\ref{eq:4-vector-for-HES-along-y-axis}).

\subsection{Hyperfine interactions for the helical edge states along the $y$-axis}

The HF interactions are now readily found by inserting the envelope HESs along the $y$-axis (\ref{eq:HES-along-y-axis}) into the general HF interaction formula (\ref{eq:HF-for-any-nanostructure}) for any structure in a HgTe QW.  Using Eq.(\ref{eq:H-HF-contact-BHZ-model}) the contact HF interaction becomes
\begin{align}
\mathcal{H}^{(y)}_{\textrm{HF},1}
=& 
\frac{1}{2\h}
\frac{B-D}{2B} 
\sum_{k_yk_y'}
\sum_n
\frac{e^{i(k_y'-k_y)\mathcal{Y}_n}}{L_y}
A_{S,j_n}(\mathcal{Z}_n)
\label{eq:HF1-HES-basis-along-y-axis-general}
\\
&{\times}\bigg[
\Lambda^{(\mathcal{X}_n)}_{k_y,k_y'}
I_{z,n}
c_{k_yu}^\dag c_{k_y'u}^{}
-
\Lambda^{(\mathcal{X}_n)}_{-k_y,-k_y'}
I_{z,n}
c_{k_yd}^\dag c_{k_y'd}^{}
\nonumber\\
&\hspace{-1mm}
-
\Lambda^{(\mathcal{X}_n)}_{k_y,-k_y'}
I_{-,n}
c_{k_yu}^\dag c_{k_y'd}^{}
-
\Lambda^{(\mathcal{X}_n)}_{-k_y,k_y'}
I_{+,n}
c_{k_yd}^\dag c_{k_y'u}^{}
\bigg],
\nonumber
\end{align}
where $c_{k_y\varsigma}^\dag$ ($c_{k_y\varsigma}^{}$) are the creation (annihilation) operators for the HESs $\Psi^{\varsigma}_{y,k_y}(x,y)$ (\ref{eq:HES-along-y-axis}). We have emphasized in the notation that $\mathcal{H}^{(y)}_{\textrm{HF},1}$ is for HESs along the $y$-axis. The product of the transverse parts of the HESs at the nuclear spin $n$ is introduced as
\begin{align}
\Lambda^{(\mathcal{X}_n)}_{k_y,k_y'}
\equiv
g^\ast_{k_y}(\mathcal{X}_n)g^{}_{k_y'}(\mathcal{X}_n),
\end{align}
and includes the only dependence of $\mathcal{X}_n$ in $\mathcal{H}^{(y)}_{\textrm{HF},1}$. Here we see that the contact HF interactions can produce transitions between the HESs $\Psi^{u}_{y,k_y}(x,y)$ and $\Psi^{d}_{y,k'_y}(x,y)$ at the expense of a change in a nuclear spin state. In particular, \emph{elastic} transitions within the Kramers pair $\Psi^{u}_{y,k_y}(x,y)$ and $\Psi^{d}_{y,-k_y}(x,y)$ are possible. This is just as if the HESs were spin-$1/2$ as used e.g.~in Refs.~\onlinecite{Lunde-Platero-PRB-2012,Maestro-Hyart-Rosenow-PRB-2013,Tanaka-Furusaki-Matveev-PRL-2011}. Hence, from the point of view of the electrons in the HESs the time reversal symmetry is broken. Of course, the composed system of electrons \emph{and} nuclear spins is time-reversal invariant, since any system can be made time-reversal invariant by expanding it sufficiently.\cite{Sakurai-modern-BOOK} 

The HF interaction due to the $P$-like states, $\mathcal{H}^{(y)}_{\textrm{HF},P}=\mathcal{H}^{(y)}_{\textrm{HF},2}+\mathcal{H}^{(y)}_{\textrm{HF},3}$, is similarly found by inserting $\Psi^{\varsigma}_{y,k_y}$ (\ref{eq:HES-along-y-axis}) into Eq.(\ref{eq:HF-for-any-nanostructure}) and using $\tilde{H}_{\textrm{HF},P}$ (\ref{eq:H-HF-P-BHZ-model}), i.e.
\begin{widetext}
\begin{align}
&\mathcal{H}^{(y)}_{\textrm{HF},P}
=
\frac{2}{15\h}
\sum_n 
\sum_{k_y,k_y'}
\frac{e^{i(k_y'-k_y)\mathcal{Y}_n}}{L_y}
\Bigg\{
-2\frac{B-D}{B}
A_{P,j_n}^{EE}
\Big[
\Lambda^{(\mathcal{X}_n)}_{k_y,-k_y'}
I_{-,n}
c_{k_yu}^\dag c_{k_y'd}^{}
+
\Lambda^{(\mathcal{X}_n)}_{-k_y,k_y'}
I_{+,n}
c_{k_yd}^\dag c_{k_y'u}^{}
\Big]
\label{eq:H-HF23-ES-basis-general-form}
\\
&\hspace{4mm}
+
\left[
{-}\frac{2\sqrt{3}A\sqrt{B^2{-}D^2}}{|A|B}
\textrm{Im}(A_{P,j_n}^{EH})
I_{x,n}
+
\frac{(B{-}D)
A_{P,j_n}^{EE}
+3(B{+}D)
A_{P,j_n}^{HH}
}{B}
I_{z,n}
\right]
\bigg[
\Lambda^{(\mathcal{X}_n)}_{k_y,k_y'}
c_{k_yu}^\dag c_{k_y'u}^{}
-
\Lambda^{(\mathcal{X}_n)}_{-k_y,-k_y'}
c_{k_yd}^\dag c_{k_y'd}^{}
\bigg]
\Bigg\},
\nonumber
\end{align}
\end{widetext}
where the dependence on $\mathcal{Z}_n$ is inside the HF couplings $A_{P,j_n}^{XY}$ Eq.(\ref{eq:position-dependent-P-HF-coulings-def}). Here we used the rewritings $i[I_{+,n}A_{P,j_n}^{EH}-I_{-,n}A_{P,j_n}^{HE}]=-2I_{x,n}\textrm{Im}(A_{P,j_n}^{EH})$ and $i[I_{+,n}A_{P,j_n}^{HE}-I_{-,n}A_{P,j_n}^{EH}]=+2I_{x,n}\textrm{Im}(A_{P,j_n}^{EH})$, which build on the fact that $A_{P,j_n}^{EH}=i\textrm{Im}(A_{P,j_n}^{EH})$ due to the phase conventions of $f_{E\Gamma_8}$ as purely imaginary and $f_H$ as real. This HF interaction also permits transitions between the two HESs -- especially within the Kramers pair -- just as the contact HF interaction (\ref{eq:HF1-HES-basis-along-y-axis-general}).  The terms $\propto c_{k\varsigma}^\dag c_{k'\varsigma}^{}$ in the HF interaction (\ref{eq:H-HF23-ES-basis-general-form}) affect transitions within a single HES. These are more unusual than their counterparts in the contact HF interaction (\ref{eq:HF1-HES-basis-along-y-axis-general}), since they do not only contain terms involving $I_{z,n}$, but also $I_{x,n}$. Hence, the HF interaction (\ref{eq:H-HF23-ES-basis-general-form}) due to the $P$-like states has terms like a $I_xS_z$ coupling, which are not present in e.g. a Heisenberg model. These terms $\propto I_{x,n} c_{k\varsigma}^\dag c_{k'\varsigma}^{}$ stem from the fact that the HF interactions due to the $P$-like states Eq.(\ref{eq:H-HF-both-P-like-BHZ-model}) couple the states $|H\pm\rangle$ and $|E\pm\rangle$ \emph{within} a single time-reversed block of $H_0$. 

In order to shine more light on the form of the HF interactions (\ref{eq:HF1-HES-basis-along-y-axis-general}) and (\ref{eq:H-HF23-ES-basis-general-form}), they are given in Appendix \ref{app:HF-in-spin-operators} in terms of non-diagonal edge state spin operators using the picture of spin-1/2 HESs.

\subsection{Hyperfine interactions for helical edge states along the $x$-axis: Curious differences}

The HF interactions presented above are for HESs running along the $y$\emph{-axis}. Now, we find various interesting differences in the HF interactions for HESs running along the $x$\emph{-axis}. 

The HESs are found in the same way as in sec. \ref{sec:HES}. The only difference is that we consider the HESs localized near a boundary given by the $x$-axis instead of the $y$-axis, i.e. we study the semi-infinite half plane defined by $y>0$. The HESs along the $x$-axis are given by 
\begin{align}
\label{eq:HES-along-x-axis}
\Psi^{\varsigma}_{x,k_x}(x,y)&=
\frac{1}{\sqrt{L_x}}
e^{ik_xx}
\mathfrak{g}^{\varsigma}_{k_x}(y)
\chi^{\varsigma}_{x}
\quad\textrm{for}\quad
\varsigma=u,d.  
\end{align}
Here $\mathfrak{g}^{u}_{k_x}(y)=g_{-k_x}(y)$ and $\mathfrak{g}^{d}_{k_x}(y)=g_{+k_x}(y)$ in terms of $g_k$ in Eq.(\ref{eq:transverse-ES-wave-function-including-lenghts}) and the spinor parts are
\begin{align}
\chi_{x}^{u}=
\mathfrak{n}
\left(
\begin{array}{c}
 \frac{A}{|A|}\\
 \frac{\sqrt{B^2-D^2}}{B-D}\\
 0\\
 0
\end{array}
\right),
\quad
\chi_{x}^{d}=
\mathfrak{n}
\left(
\begin{array}{c}
 0\\
 0\\
 \frac{A}{|A|}\\
 \frac{\sqrt{B^2-D^2}}{B-D}
\end{array}
\right),
\label{eq:4-vector-for-HES-along-x-axis}
\end{align}
i.e. \emph{the imaginary unit} $i$ \emph{does not appear} in the $E\pm$ components of the spinors as for the HESs along the $y$-axis, see Eq.(\ref{eq:4-vector-for-HES-along-y-axis}). \emph{This is the mathematical origin of the differences between the HF interactions for the HESs in the two directions}. These HESs also appear in Kramers pairs ($\Psi^{u}_{x,k_x}$ and $\Psi^{d}_{x,-k_x}$) and $\Psi^{u}_{x,k_x}$ ($\Psi^{d}_{x,k_x}$) is referred to as spin-up (spin-down). The spin-up HES $\Psi^{u}_{x,k_x}$ has negative velocity such that $E^u_{k_x}=E_0-\h v_0 k_x$, while the spin-down HES has positive velocity, i.e. $E^d_{k_x}=E_0+\h v_0 k_x$. Hence, the velocities of the HESs along the $x$ and $y$ axes have opposite signs, such that spin-$\varsigma$ always travels the same way along the boundary, see Fig. \ref{fig:HES-corner}. Therefore, it is natural that $k_x$ has to be exchanged by $-k_y$ to connect the HESs in the two perpendicular directions.

By inserting the HESs along the $x$-axis Eq.(\ref{eq:HES-along-x-axis}) into Eq.(\ref{eq:HF-for-any-nanostructure}), the contact HF interaction becomes 
\begin{align}
\mathcal{H}^{(x)}_{\textrm{HF},1}
=& 
\frac{1}{2\h}
\frac{B-D}{2B} 
\sum_{k_xk_x'}
\sum_n
\frac{e^{i(k_x'-k_x)\mathcal{X}_n}}{L_x}
A_{S,j_n}(\mathcal{Z}_n)
\label{eq:HF1-HES-basis-along-x-axis-general}
\\
&{\times}\bigg[
\Lambda^{(\mathcal{Y}_n)}_{-k_x,-k_x'}
I_{z,n}
c_{k_xu}^\dag c_{k_x'u}^{}
-
\Lambda^{(\mathcal{Y}_n)}_{k_x,k_x'}
I_{z,n}
c_{k_xd}^\dag c_{k_x'd}^{}
\nonumber\\
&\hspace{-2mm}
+
\Lambda^{(\mathcal{Y}_n)}_{-k_x,k_x'}
I_{-,n}
c_{k_xu}^\dag c_{k_x'd}^{}
+
\Lambda^{(\mathcal{Y}_n)}_{k_x,-k_x'}
I_{+,n}
c_{k_xd}^\dag c_{k_x'u}^{}
\bigg],
\nonumber
\end{align}
where $\Lambda^{(\mathcal{Y}_n)}_{k_x,k_x'}=g^\ast_{k_x}(\mathcal{Y}_n)g^{}_{k_x'}(\mathcal{Y}_n)$. Interestingly, the sign of the terms producing inter HES transitions is opposite to the one in $\mathcal{H}^{(y)}_{\textrm{HF},1}$ (\ref{eq:HF1-HES-basis-along-y-axis-general}). This difference stems from the imaginary unit $i$ in $\chi_y^{\varsigma}$ (\ref{eq:4-vector-for-HES-along-y-axis}), which is absent in $\chi_x^{\varsigma}$ (\ref{eq:4-vector-for-HES-along-x-axis}). Moreover, the sign of $k_x$ and $k_y$ is opposite in the $\Lambda$ functions for $\mathcal{H}^{(x)}_{\textrm{HF},1}$ (\ref{eq:HF1-HES-basis-along-x-axis-general}) and $\mathcal{H}^{(y)}_{\textrm{HF},1}$ (\ref{eq:HF1-HES-basis-along-y-axis-general}), respectively. This is natural in order to maintain the propagation direction of the HES-spin $\varsigma=u,d$, see Fig. \ref{fig:HES-corner}.  

The HF interaction due to the $P$-like states for the HESs at the $x$-axis becomes
\begin{widetext}
\begin{align}
&\mathcal{H}^{(x)}_{\textrm{HF},P}
=
\frac{2}{15\h}
\sum_n 
\sum_{k_x,k_x'}
\frac{e^{i(k_x'-k_x)\mathcal{X}_n}}{L_x}
\Bigg\{
+2\frac{B-D}{B}
A_{P,j_n}^{EE}
\Big[
\Lambda^{(\mathcal{Y}_n)}_{-k_x,k_x'}
I_{-,n}
c_{k_xu}^\dag c_{k_x'd}^{}
+
\Lambda^{(\mathcal{Y}_n)}_{k_x,-k_x'}
I_{+,n}
c_{k_xd}^\dag c_{k_x'u}^{}
\Big]
\label{eq:H-HF23-ES-basis-general-form-x-axis}
\\
&\hspace{5mm}
+
\left[
-\frac{2\sqrt{3}A\sqrt{B^2{-}D^2}}{|A|B}
\textrm{Im}(A_{P,j_n}^{EH})
I_{y,n}
+
\frac{(B{-}D)
A_{P,j_n}^{EE}
+3(B{+}D)
A_{P,j_n}^{HH}}{B}
I_{z,n}
\right]
\Big[
\Lambda^{(\mathcal{Y}_n)}_{-k_x,-k_x'}
c_{k_xu}^\dag c_{k_x'u}^{}
-
\Lambda^{(\mathcal{Y}_n)}_{k_x,k_x'}
c_{k_xd}^\dag c_{k_x'd}^{}
\Big]
\Bigg\},
\nonumber
\end{align}
\end{widetext}
where the inter HES transition terms again have an opposite overall sign compared to $\mathcal{H}^{(y)}_{\textrm{HF},P}$ (\ref{eq:H-HF23-ES-basis-general-form}). Another noteworthy difference is the exchange of the terms $I_{x,n}c_{k_y\varsigma}^\dag c_{k_y'\varsigma}^{}$ in $\mathcal{H}^{(y)}_{\textrm{HF},P}$ by $I_{y,n}c_{k_x\varsigma}^\dag c_{k_x'\varsigma}^{}$  in $\mathcal{H}^{(x)}_{\textrm{HF},P}$, i.e. \emph{intra HES transitions are coupled to the nuclear spin operators perpendicular to the propagation direction}. These differences again stem from the imaginary unit (or the lack thereof) in the spinors. Furthermore, the signs of $k_x$ and $k_y$ are again interchanged in the $\Lambda$ functions by comparing $\mathcal{H}^{(y)}_{\textrm{HF},P}$ and $\mathcal{H}^{(x)}_{\textrm{HF},P}$.

\subsection{Position averaged hyperfine interactions}\label{sec:position-average}

In HgTe about 19$\%$ of the atoms have a non-zero nuclear spin and these can be assumed to be randomly distributed. In the HF interactions for the HESs the unit-cell position of every nuclear spin is included. This information is sample dependent and often valuable insights can be found \emph{without} it. Therefore, we now consider the HF interactions averaged over the unit-cell position of the nuclear spins in analogue to impurity averaging.\cite{Flensberg-BOOK} To be specific, we focus here on the HF interactions (\ref{eq:HF1-HES-basis-along-y-axis-general}) and (\ref{eq:H-HF23-ES-basis-general-form}) for the HESs along the $y$-axis. Mathematically, the position averaged of some quantity $\mathcal{F}$ is introduced as 
\begin{align}
\overline{\mathcal{F}}\equiv
\frac{1}{\mathcal{A}^{N_s}}\!
\int_{\mathcal{A}}\!d \Ruc_{\diamond,1}\cdots
\int_{\mathcal{A}}\!d \Ruc_{\diamond,N_s} \mathcal{F}(\Ruc_1,\ldots,\Ruc_{N_s}),
\label{eq:imp-ave-def}  
\end{align}
where $N_s$ is the number of non-zero nuclear spins covered by the HESs. We only average over the positions $\Ruc_{\diamond,n}\equiv(\mathcal{X}_n,\mathcal{Z}_n)$ in the cross section area $\mathcal{A}$ of the HESs along the $y$-axis. Thereby we keep the positions $\mathcal{Y}_n$, which break translational invariance along the edge and ultimately can lead to backscattering.\cite{Tanaka-Furusaki-Matveev-PRL-2011,Lunde-Platero-PRB-2012,Eriksson-Strom-Sharma-Johannesson-PRB-2012,Maestro-Hyart-Rosenow-PRB-2013}   

Now we study the position averaged HF Hamiltonians. However, one can equally well position average at a later stage of a calculation, if it is physically relevant for a particular phenomenon, e.g. position averaging of the transition rates.\cite{Lunde-Platero-PRB-2012} Using the normalization of $f_{E\Gamma_6}$ in Appendix \ref{App:normalization}, the position averaged contact HF interaction (\ref{eq:HF1-HES-basis-along-y-axis-general}) becomes
\begin{align}
\overline{\mathcal{H}^{(y)}_{\textrm{HF},1}}
=& 
\frac{1}{4\h}
\frac{B-D}{2B} 
\sum_{k_yk_y'}
\sum_n
e^{i(k_y'-k_y)\mathcal{Y}_n}
\frac{A^{\textrm{Atomic}}_{S,j_n}}{N}
\label{eq:HF1-HES-basis-along-y-axis-position-averged}\\
&{\times}\bigg[
\overline{\Lambda_{k_y,k_y'}}
I_{z,n}
c_{k_yu}^\dag c_{k_y'u}^{}
-
\overline{\Lambda_{-k_y,-k_y'}}
I_{z,n}
c_{k_yd}^\dag c_{k_y'd}^{}
\nonumber\\
&\hspace{-1mm}
-
\overline{\Lambda_{k_y,-k_y'}}
I_{-,n}
c_{k_yu}^\dag c_{k_y'd}^{}
-
\overline{\Lambda_{-k_y,k_y'}}
I_{+,n}
c_{k_yd}^\dag c_{k_y'u}^{}
\bigg],
\nonumber
\end{align}
where the cross section area $\mathcal{A}$ is given in terms of the QW thickness $W_z$ and the HES width $W_x$ as $\mathcal{A}=W_xW_z$, such that $\int_{\mathcal{A}}d \Ruc_{\diamond,i}=1$. Here $W_x$ is on the order of a few decay lengths $\lambda_2^{-1}$ and the number of atoms covered by the HESs is $N\equiv (L\mathcal{A})/v_a$. We observe that the position dependent HF coupling $A^{}_{S,j_n}(\mathcal{Z}_n)$ (\ref{eq:def-HF-position-dependent-contact-coupling}) is replaced by a homogenous HF coupling $A^{\textrm{Atomic}}_{S,j_n}/N$ due to the position averaging as in the case of quantum dots.\cite{Khaetskii-Loss-Glazman-PRB-2003} The position average of the product of transverse functions, $\overline{\Lambda_{k,k'}}= \int_{0}^{W_x}d\mathcal{X}_n g^\ast_{k}(\mathcal{X}_n)g^{}_{k'}(\mathcal{X}_n)$, is now independent of the positions $\mathcal{X}_n$. It can be well approximated by replacing $W_x$ by $\infty$ in the upper limit, which gives
\begin{align}
\overline{\Lambda_{k_y,k_y'}}=& 
\sqrt{\frac{2\lambda_{1}\lambda_{2} (\lambda_{1}+\lambda_{2})}{(\lambda_{1}-\lambda_{2})^2}} 
\sqrt{\frac{2\lambda'_{1}\lambda'_{2} (\lambda'_{1}+\lambda'_{2})}{(\lambda'_{1}-\lambda'_{2})^2}} 
\label{eq:Lambda-averaged-full-form}
\\
&\times
\left[\frac{1}{\lambda'_{1}+\lambda_{1}}
-\frac{1}{\lambda'_{2}+\lambda_{1}}
-\frac{1}{\lambda'_{1}+\lambda_{2}}
+\frac{1}{\lambda'_{2}+\lambda_{2}}
\right],
\nonumber
\end{align}
where $\lambda_{i}$ and $\lambda'_{i}$ depends on $k_y$ and $k_y'$, respectively. It is evident that $\overline{\Lambda_{k_y,k_y'}}=\overline{\Lambda_{k_y',k_y}}$, since $g_{k_y}$ in Eq.(\ref{eq:transverse-ES-wave-function}) is real. Moreover, $\overline{\Lambda_{k_y,k_y}}=1$ due to the normalization of $g_{k_y}$.  Furthermore, in the particle-hole symmetric limit $D=0$, we have $g_{-k_y}(x)=g_{k_y}(x)$ such that  $\overline{\Lambda_{k_y,k_y'}}^{(D=0)}=\overline{\Lambda_{-k_y,-k_y'}}^{(D=0)}=\overline{\Lambda_{k_y,-k_y'}}^{(D=0)}=\overline{\Lambda_{-k_y,k_y'}}^{(D=0)}$. Hence, the position averaged contact HF interaction (\ref{eq:HF1-HES-basis-along-y-axis-position-averged}) becomes \emph{isotropic} in the particle-hole symmetric limit. Since the BHZ model is valid only close to the $\Gamma$ point, we expand $\overline{\Lambda_{k_y,k_y'}}$ to lowest order in $k_y$ and $k_y'$ for $D\neq0$, i.e. $\overline{\Lambda_{k_y,k_y'}}\simeq\eta_{k_y,k_y'}$, where 
\begin{align}
\eta_{k_y,k_y'}=
1-
\frac{D^2 \big[A^2 B{+}2 (B^2{-}D^2) M_0\big]}{8 B M_0^2(B^2-D^2)}
(k_y-k_y')^2.
\label{eq:eta-2nd-order-in-k}
\end{align}
Hence, the lowest order expansion fulfills $\eta_{k_y,k_y'}=\eta_{-k_y,-k_y'}$ and $\eta_{-k_y,k_y'}=\eta_{k_y,-k_y'}$ such that the position averaged contact HF interaction (\ref{eq:HF1-HES-basis-along-y-axis-position-averged}) simplifies to
\begin{align}
\overline{\mathcal{H}^{(y)}_{\textrm{HF},1}}
\simeq& 
\frac{1}{4\h}
\frac{B-D}{2B} 
\sum_{k_yk_y'}
\sum_n
e^{i(k_y'-k_y)\mathcal{Y}_n}
\frac{A^{\textrm{Atomic}}_{S,j_n}}{N}
\label{eq:HF1-HES-basis-along-y-axis-position-averged-low-k}\\
&{\times}\bigg[
\eta_{k_y,k_y'}
I_{z,n}
(c_{k_yu}^\dag c_{k_y'u}^{}
-
c_{k_yd}^\dag c_{k_y'd}^{})
\nonumber\\
&\hspace{4mm}
-\eta_{k_y,-k_y'}
(I_{-,n}
c_{k_yu}^\dag c_{k_y'd}^{}
+
I_{+,n}
c_{k_yd}^\dag c_{k_y'u}^{})
\bigg],
\nonumber
\end{align}
to lowest order in $k_y$ and $k_y'$. In this limit, $\overline{\mathcal{H}^{(y)}_{\textrm{HF},1}}$ therefore has uniaxial anisotropy.

The position averaged HF interaction due to the $P$-like states in Eq.(\ref{eq:H-HF23-ES-basis-general-form}) becomes
\begin{align}
\overline{\mathcal{H}^{(y)}_{\textrm{HF},P}}
\simeq&
\frac{2}{15\h}
\sum_n 
\sum_{k_y,k_y'}
e^{i(k_y'-k_y)\mathcal{Y}_n}
\frac{A_{P,j_n}^{\textrm{Atomic}}}{N}
\label{eq:HFP-HES-basis-along-y-axis-position-averged-low-k}
\\
&\hspace{-2mm}
{\times}\Bigg\{
\eta^{}_{k_y,k_y'}
\frac{7B+5D}{2B}
I_{z,n}
\big[
c_{k_yu}^\dag c_{k_y'u}^{}
-
c_{k_yd}^\dag c_{k_y'd}^{}
\big]
\nonumber\\
&\hspace{0mm}
{-}
\frac{B{-}D}{B}
\eta^{}_{k_y,-k_y'}\!
\Big[I_{-,n}
c_{k_yu}^\dag c_{k_y'd}^{}
+
I_{+,n}
c_{k_yd}^\dag c_{k_y'u}^{}
\Big]
\Bigg\}
\nonumber
\end{align}
by using the expansion $\overline{\Lambda_{k_y,k_y'}}\simeq\eta_{k_y,k_y'}$ and the normalization conditions for $f_H$ and $f_{E\Gamma_8}$ (see Appendix \ref{App:normalization}). Interestingly, the terms in $\mathcal{H}^{(y)}_{\textrm{HF},P}$ (\ref{eq:H-HF23-ES-basis-general-form}) coupling $I_{x,n}$ and  $c_{k_y\varsigma}^\dag c_{k_y'\varsigma}^{}$ \emph{vanish} in the position averaging, since $f_{H}(z)$ is even and $f_{E\Gamma_8}(z)$ is odd\cite{Bernevig-Zhang-Science-2006,Hankiewicz-NJP-2010} such that $\overline{A_{P,j_n}^{EH}}\propto\int d\mathcal{Z}_nf_{E\Gamma_8}^\ast(\mathcal{Z}_n)f_{H}(\mathcal{Z}_n)=0$. Furthermore, even in the particle-hole symmetric limit $D=0$, $\overline{\mathcal{H}^{(y)}_{\textrm{HF},P}}$ is not isotropic in contrast to the contact HF interaction.

\begin{table}
  \begin{tabular}{ l | c | c | c | c | }
     & $^{199}$Hg & $^{201}$Hg & $^{123}$Te & $^{125}$Te \\ 
    \hline
     \multirow{2}{*}{$A^{z}_{j_n}$ [$\mu$eV]} & 1.1 & -0.38 & -12 & -14 \\
     & (77$\%$) & (75$\%$) & (73$\%$) & (73$\%$) \\ 
    \hline
     \multirow{2}{*}{$A^{\bot}_{j_n}$ [$\mu$eV]} & 0.30 & -0.11 & -3.5 & -4.2 \\
     & (14$\%$) & (12$\%$) & (12$\%$) & (12$\%$) \\  
    \hline
  \end{tabular}
\caption{Estimates of the effective HF couplings (\ref{eq:eff-HF-couplings-fo-HES}) for a pair of HESs. In parentheses we give the percentage of the HF coupling stemming from the HF hamiltonians due to $P$-like states, e.g. $4[(B-D)/(15B)]A_{P,j_n}^{\textrm{Atomic}}/A^{\bot}_{j_n}$. Here we use the atomic HF couplings in table \ref{Table:atomic-S-HF-couplings} and the BHZ model parameters\cite{Qi-Zhang-review-RMP-2010,footnote-parameters-for-BHZ-model} $B$ and $D$ only for a 70\AA\ thick QW.}
\label{Table:eff-HF-couplings-for-HES}
\end{table}

The total position averaged HF interaction $\overline{\mathcal{H}^{(y)}_{\textrm{HF}}}=\overline{\mathcal{H}^{(y)}_{\textrm{HF},1}}+\overline{\mathcal{H}^{(y)}_{\textrm{HF},P}}$ in the small wave-vector limit is now found from Eqs.(\ref{eq:HF1-HES-basis-along-y-axis-position-averged-low-k},\ref{eq:HFP-HES-basis-along-y-axis-position-averged-low-k}) to be 
\begin{align}
\overline{\mathcal{H}^{(y)}_{\textrm{HF}}}
\simeq& 
\frac{1}{2\h}
\sum_{n,k_yk_y'}
e^{i(k_y'-k_y)\mathcal{Y}_n}
\label{eq:HF-total-HES-basis-along-y-axis-position-averged-low-k}\\
&\times
\bigg[
\frac{A^{z}_{j_n}}{N}
\eta^{}_{k_y,k_y'}
I_{z,n}
\big(c_{k_yu}^\dag c_{k_y'u}^{}
-
c_{k_yd}^\dag c_{k_y'd}^{}\big)
\nonumber\\
&\hspace{2mm}
-
\frac{A^{\bot}_{j_n}}{N}
\eta^{}_{k_y,-k_y'}
\big(I_{-,n}
c_{k_yu}^\dag c_{k_y'd}^{}
+
I_{+,n}
c_{k_yd}^\dag c_{k_y'u}^{}
\big)
\bigg],
\nonumber
\end{align}
where effective HF couplings were introduced as 
\begin{subequations}
\label{eq:eff-HF-couplings-fo-HES}
\begin{align}
A^{z}_{j_n}&\equiv
\frac{B-D}{4B}
A^{\textrm{Atomic}}_{S,j_n}
+
\frac{4}{15}\frac{7B+5D}{2B}
A^{\textrm{Atomic}}_{P,j_n},\\
A^{\bot}_{j_n}&\equiv
\frac{B-D}{B}
\bigg[
\frac{1}{4}
A^{\textrm{Atomic}}_{S,j_n}
+
\frac{4}{15}
A^{\textrm{Atomic}}_{P,j_n}
\bigg].
\end{align}
\end{subequations}
Hence, the total position averaged HF interaction $\overline{\mathcal{H}^{(y)}_{\textrm{HF}}}$ has uniaxial anisotropy. Estimates of $A^{z}_{j_n}$ and $A^{\bot}_{j_n}$ are given in table \ref{Table:eff-HF-couplings-for-HES}. Remarkably, the part of the effective HF couplings due to the $P$-like states \emph{dominates} for the coupling $A^{z}_{j_n}$, but \emph{not} for $A^{\bot}_{j_n}$. One reason is that the HESs have their main contribution on the $H$ states compared to the $E$ states, since $(\chi_y^u)^T\simeq(-i0.36,-0.93,0,0)$ for a 70\AA\ thick QW.\cite{footnote-parameters-for-BHZ-model} Moreover, not only the $H$ states are $P$-like states, but also partly the $E$ states, see Eq.(\ref{eq:basis-states-mellem2}).  

We remark that the position averaged HF interactions for the HESs \emph{along the} $x$-\emph{axis} Eqs.(\ref{eq:HF1-HES-basis-along-x-axis-general},\ref{eq:H-HF23-ES-basis-general-form-x-axis}) follow along the same lines. The only difference in the total HF interaction in Eq.(\ref{eq:HF-total-HES-basis-along-y-axis-position-averged-low-k}) is an opposite sign of the inter HES transition terms (apart from the replacements $\mathcal{Y}_n\rightarrow\mathcal{X}_n$ and $k_y\rightarrow k_x$).

For typical parameters,\cite{footnote-estimate-number-of-atoms-within-HES} the number of atoms covered by the HESs is about $N\sim10^{7}$ per $\mu$m edge, where about 19$\%$ of these atoms have a non-zero nuclear spin.

\section{Discussion, summary and outlook}

In this paper, we have provided benchmark results within the BHZ model for the form and magnitude of (i) the contact HF interaction in Eq.(\ref{eq:H-HF-contact-BHZ-model}), (ii) the dipole-dipole like HF interaction in Eq.(\ref{eq:H-HF-dipole-BHZ-model}) and (iii) the coupling of the electrons orbital momentum to the nuclear spin in Eq.(\ref{eq:H-HF-3-BHZ-model}). 

All the HF interactions couple the time-reversed blocks of the BHZ Hamiltonian (\ref{eq:H-BHZ-model-matrix1}) -- just as the Rashba spin-orbit coupling\cite{Hankiewicz-NJP-2010}  and the bulk inversion asymmetry terms.\cite{Konig-JPSJ-review-2008}  However, in contrast to the Rashba and bulk inversion asymmetry terms, the HF interactions break time-reversal symmetry from the electronic point of view. Therefore, the HF interactions couple directly the Kramers pair of counterpropagating HESs of opposite wave numbers ($k$ and $-k$), and thereby open for \emph{elastic} backscattering. In contrast, the Rashba spin-orbit interaction combined with other scattering mechanisms can only couple the HESs \emph{inelastically}.\cite{Schmidt-Rachel-Oppen-Glazman-PRL-2012,Crepin-et-al-PRB-2012,Budich-Dolcini-Recher-Trauzettel-PRL-2012,Eriksson-Strom-Sharma-Johannesson-PRB-2012} Hence, our careful microscopic modelling of the HF interactions confirms that elastic backscattering spin-flip processes indeed are present as correctly anticipated on physical grounds in previous works on the interaction between HESs (modelled as spin-$1/2$) and  one or more fixed magnetic moments.\cite{Lunde-Platero-PRB-2012,Maestro-Hyart-Rosenow-PRB-2013,Tanaka-Furusaki-Matveev-PRL-2011,Eriksson-Strom-Sharma-Johannesson-PRB-2012,Eriksson-PRB-2013}

Furthermore, we estimated the atomic HF constants relevant for a HgTe QW, see table \ref{Table:atomic-S-HF-couplings}. These estimates are generally smaller by an order of magnitude or so compared to similar estimates for GaAs by Fischer \emph{et al.}\cite{Fischer-Coish-Bulaev-Loss-PRB-2008,footnote-def-of-HF-couplings} This is natural, since heavier elements often have lower HF couplings due to their higher principal quantum number of the outermost electron [see e.g. Eqs.(\ref{eq:S-HF-coupling-appendix},\ref{eq:P-HF-coupling-appendix})]. As a consequence, the typical time for polarizing the nuclear spins by a current through the HESs\cite{Lunde-Platero-PRB-2012} of a HgTe QW is increased to hours  or days
compared to seconds for a GaAs QW in the quantum hall regime.\cite{Wald-Kouwenhoven-McEuen-et-al-PRL-1994}

From the HF Hamiltonians within the BHZ model, we derived a general formula (\ref{eq:HF-for-any-nanostructure}) for the HF interactions for any nanostructure in a HgTe QW. The input of this formula is the envelope function of the given structure, where the effects of bulk inversion asymmetry,\cite{Konig-JPSJ-review-2008} Rashba spin-orbit coupling\cite{Hankiewicz-NJP-2010} or magnetic fields\cite{Scharf-Matos-Abiague-Fabian-PRB-2012} can be included. From this formula, we found the HF interactions for a pair of HESs. Interestingly, the HF Hamiltonians depend on the orientation of the boundary at which the HESs propagate: The sign of the terms creating inter HES transitions is opposite for perpendicular boundaries. This has not been considered previously in works on HESs coupled to fixed spins.\cite{Lunde-Platero-PRB-2012,Maestro-Hyart-Rosenow-PRB-2013,Tanaka-Furusaki-Matveev-PRL-2011,Eriksson-Strom-Sharma-Johannesson-PRB-2012,Eriksson-PRB-2013} On the level of transition rates between the HESs,\cite{Lunde-Platero-PRB-2012,Maestro-Hyart-Rosenow-PRB-2013,Tanaka-Furusaki-Matveev-PRL-2011} such a difference is less important, since the rates are proportional to the HF matrix elements squared. However, this sign might play a role for more delicate phenomena such as Kondo physics\cite{Eriksson-Strom-Sharma-Johannesson-PRB-2012,Eriksson-PRB-2013} or for HESs circulating one or more fixed spins.

We also found that the HF interactions due to the $P$-like states couple the \emph{intra} HES transitions to \emph{both} nuclear spin components perpendicular to the propagation direction of the HESs, see Eqs.(\ref{eq:H-HF23-ES-basis-general-form},\ref{eq:H-HF23-ES-basis-general-form-x-axis}). The unusual terms coupling $I_{x,n}$ ($I_{y,n}$) to the intra HES transitions for propagation along the $y$-axis ($x$-axis) were not included in previous studies.\cite{Lunde-Platero-PRB-2012,Maestro-Hyart-Rosenow-PRB-2013,Tanaka-Furusaki-Matveev-PRL-2011,Eriksson-Strom-Sharma-Johannesson-PRB-2012,Eriksson-PRB-2013} These terms might complicate the nature of nuclear spin polarization and its associated Overhauser effective magnetic field\cite{Overhauser-PR-1953} in a non-trivial way. For instance, this could affect the spin-orbit interaction induced backscattering processes between the HESs in the presence of a finite Overhauser field discussed in Ref. \onlinecite{Maestro-Hyart-Rosenow-PRB-2013}. 

Finally, we averaged over the positions of the nuclear spins to remove the sample dependent information. This revealed that the total HF Hamiltonian is quite generally anisotropic and, moreover, that the contribution due to $P$-like states can dominate over the contact HF contribution, see table \ref{Table:eff-HF-couplings-for-HES} and Eq.(\ref{eq:HF-total-HES-basis-along-y-axis-position-averged-low-k}). Therefore, it can be important to include the HF interactions (\ref{eq:HF2-of-nuclei-nr-n},\ref{eq:HF3-of-nuclei-nr-n}) relevant for $P$-like states for the HESs.  Moreover, we found that the coupling of $I_{x,n}$ ($I_{y,n}$) to the intra HES transitions for propagation along the $y$-axis ($x$-axis) vanishes in the position averaging of the HF Hamiltonians. In this sense, these couplings are somewhat fragile compared to the usual coupling of $I_{z,n}$ to the intra HES transitions. On the other hand, position averaging at a later stage of a calculation might allow interesting effects from these unusual terms to survive.    

In passing, we remark that the nuclear spins can open a very small energy gap in the HES spectrum. This can be shown by averaging out \emph{all} spacial directions of the nuclear spin positions in the total HF interaction. Treating the nuclear spins as a semi-classical field of zero mean value,\cite{Merkulov-Efros-Rosen-PRB-2002,Erlingsson-Nazarov-PRB-2002} the energy gap becomes proportional to the in-plane field. The ensemble averaged energy gap\cite{footnote-energy-gab-details} is proportional to $N^{-1/2}$ and estimated to be on the order of $10^{-4}\mu$eV for a micron sized edge, which seems out of the current experimental range.

\section{Acknowledgments}
We are especially grateful to Jan Fischer and Dietrich Rothe for helpful correspondence on their work in Refs.[\onlinecite{Fischer-Loss-PRL-2010,Fischer-Trauzettel-Loss-PRB-2009,Fischer-Trif-Coish-Loss-Solid-state-comm-2009,Fischer-Coish-Bulaev-Loss-PRB-2008}] and Ref.[\onlinecite{Hankiewicz-NJP-2010}], respectively. We also thank Laurens Molenkamp, Bj\"{o}rn Trauzettel, Andrzej K\c{e}dziorski, Jens Paaske and Karsten Flensberg for useful discussions. Both AML and GP are supported by Grant No. MAT2011-24331 and by the ITN Grant 234970 (EU). AML acknowledges the Juan de la Cierva program (MICINN) and Grant No.~FIS2009-07277 and the Carlsberg Foundation.  Furthermore, we acknowledge FIS2010-22438-E (Spanish National Network for Physics of Out-of-Equilibrium Systems).

\appendix

\section{On the BHZ model states}\label{App:TR-af-basis-states}

This appendix describes various details of the BHZ states $|E\pm\rangle$ and $|H\pm\rangle$. In particular, the time-reversal properties and the phase conventions of the envelope functions are discussed. 

The states in the BHZ model as presented in Ref.[\onlinecite{Bernevig-Zhang-Science-2006}] are given by
\begin{subequations}
\begin{align}
|E+\rangle&=f_1(z)|\Gamma_6,+1/2\rangle+f_4(z)|\Gamma_8,+1/2\rangle,\\   
|H+\rangle&=f_3(z)|\Gamma_8,+3/2\rangle,\\   
|E-\rangle&=f_2(z)|\Gamma_6,-1/2\rangle+f_5(z)|\Gamma_8,-1/2\rangle,\\   
|H-\rangle&=f_6(z)|\Gamma_8,-3/2\rangle,
\end{align}
\end{subequations}
similar to Eq.(\ref{eq:basis-states-mellem2}), but without specifying any phase conventions for the envelope functions $f_n(z)$. The lattice periodic functions can be given as 
\begin{subequations}
\label{eq:Gamma-states-S}
\begin{align}
&|\Gamma_6,+1/2\rangle
=
|S\rangle |\!\uparrow\rangle, \\
&|\Gamma_6,-1/2\rangle
=
|S\rangle |\!\downarrow\rangle, 
\end{align}
\end{subequations}
and
\begin{subequations}
\label{eq:Gamma-states-P}
\begin{align}
&|\Gamma_8,3/2\rangle=
+\frac{1}{\sqrt{2}}\Big[|P_x\rangle+i|P_y\rangle\Big] |\!\uparrow\rangle, \\
&|\Gamma_8,1/2\rangle=
-\sqrt{\frac{2}{3}}|P_z\rangle |\!\uparrow\rangle
+\!\frac{1}{\sqrt{6}}\big[|P_x\rangle+i|P_y\rangle\big] |\!\downarrow\rangle,\\
&|\Gamma_8,-1/2\rangle=
-\sqrt{\frac{2}{3}}|P_z\rangle |\!\downarrow\rangle
{-}\frac{1}{\sqrt{6}}\Big[|P_x\rangle{-}i|P_y\rangle\Big] |\!\uparrow\rangle, \\
&|\Gamma_8,-3/2\rangle=
-\frac{1}{\sqrt{2}}\Big[|P_x\rangle-i|P_y\rangle\Big] |\!\downarrow\rangle,
\end{align}
\end{subequations}
where the Bloch amplitudes $|S\rangle$, $|P_x\rangle$, $|P_y\rangle$ and $|P_z\rangle$ transform the same way as the well-known orbitals with the same names.\cite{Fabian-review-2007,Yu-Cardona-BOOK-2001} The orbitals are connected to the spherical harmonics.\cite{Sakurai-modern-BOOK,Winkler-BOOK-2003} Thus, $|\Gamma_8,m_j\rangle$ correspond to $|j=3/2,m_j,l=1,s=1/2\rangle$ in the angular momentum representation using the total angular momentum $\Jj=\Ll+\Ss$ as a good quantum number, where $\Ss$ is the electron spin in the basis $\{\op,\ned\}$. Likewise, $|\Gamma_6,m_j\rangle$ simply corresponds to the $l=0$ state. Note that the split-off band $\Gamma_7$ with $j=1/2$ and $l=1$ is neglected in the BHZ model. Here $|S\rangle$ is chosen to be purely imaginary\cite{footnote-S-imaginary} and $|P_x\rangle$, $|P_y\rangle$ and $|P_z\rangle$ to be real.\cite{Winkler-BOOK-2003} Furthermore, we follow the convention by Bernevig \emph{et al.}\cite{Bernevig-Zhang-Science-2006} and Novik \emph{et al.}\cite{Novik-Molenkamp-PRB-2005} by using an overall opposite sign\cite{footnote-overall-sign-of-P-states} for the $\Gamma_8$ states in terms of the $P$ states in Eq.(\ref{eq:Gamma-states-P}) compared to other authors.\cite{Winkler-BOOK-2003,Katsaros-Glazman-et-al-PRL-2011} This sign change is not important for the purposes of this paper.  

Next we discuss the phase conventions for the envelope functions made in the main text. The envelope functions $f_n(z)$ ($n=1,\ldots,6$) are found from the Luttinger-Kane model at $k_x=k_y=0$ and therefore has to fulfill the following differential equations\cite{Bernevig-Zhang-Science-2006,Hankiewicz-NJP-2010}  
\begin{subequations}
\label{eq:diff-eq-for-f1f4-og-f2f5-par}
\begin{align}
Tf_{n}(z)-\sqrt{\frac{2}{3}}P_0i\p_z f_{n+3}(z)&=E_{\kk=0}f_n(z),\\
-\sqrt{\frac{2}{3}}P_0i\p_zf_{n}(z)+ W_-f_{n+3}(z)&=E_{\kk=0}f_{n+3}(z), 
\end{align}
\end{subequations}  
for $n=1,2$ only [i.e. only for the two pairs $(f_1,f_4)$ and $(f_2,f_5)$]. Similarly,\cite{Bernevig-Zhang-Science-2006,Hankiewicz-NJP-2010}   
\begin{align}
W_+f_{n}(z)&=E_{\kk=0}f_{n}(z), 
\quad \textrm{for}\ n=3,6\ \textrm{only}.
\label{eq:diff-eq-for-f3-og-f6}
\end{align}
Here we have introduced the real operators 
\begin{subequations}  
\begin{align}
T&=E_c(z)+\frac{\h^2}{2m_e}k_z(2F(z)+1)k_z,\\
W_{\pm}&=E_v(z)+\frac{\h^2}{2m_e}k_z(2\gamma_2(z)\mp\gamma_1(z))k_z,
\end{align}
\end{subequations}  
where $k_z=-i\p_z$, $E_{c(v)}$ is the conduction (valence) band edge, $m_e$ the bare electron mass, $\gamma_{1,2}$ are the Luttinger parameters,\cite{Luttinger-PR-1956} and $F(z)$ is a real function including the remote $\Gamma_5$ bands perturbatively.\cite{Hankiewicz-NJP-2010} The parameters $\gamma_{1,2}$, $F$ and $E_{c,v}$ are different in the HgTe and CdTe layers of the heterostructure, which leads to the  $z$-dependence. The solution of these equations will also give the energy for that particular solution (energy band) $E_{\kk=0}$ at $\kk=(0,0)$. From Eq.(\ref{eq:diff-eq-for-f3-og-f6}) it follows that we can choose $f_3(z)=f_6(z)$, which is simply denoted as $f_{H}(z)$ in the main text. Furthermore, Eqs.(\ref{eq:diff-eq-for-f1f4-og-f2f5-par}) allow us to choose $f_1(z)=f_2(z)$ and $f_4(z)=f_5(z)$, which are called $f_{E\Gamma_6}(z)$ and $f_{E\Gamma_8}(z)$, respectively, in the main text. By comparison of Eqs.(\ref{eq:diff-eq-for-f1f4-og-f2f5-par}) and their complex conjugates, it follows that we can choose $f_1(z)$ real and $f_4^{}(z)$ purely imaginary as in Ref.[\onlinecite{Hankiewicz-NJP-2010}]. 

Now we turn our attention to the time-reversal properties of the states $|E\pm\rangle$ and $|H\pm\rangle$. The time-reversal operator $\Theta$ is defined up to an arbitrary phase factor. Here we use $\Theta=-i\sigma_yK$, where $K$ is the complex conjugation operator and $\sigma_y$ a Pauli matrix in electron spin-space. The time-reversal operator $\Theta$ acts differently in different bases (due to the complex conjugation), so one should stick to the same basis through out a calculation.\cite{Sakurai-modern-BOOK} The $|\Gamma_i,m_j\rangle$ states under the time-reversal operator follow from Eqs.(\ref{eq:Gamma-states-S}) and (\ref{eq:Gamma-states-P}) by using that the $P$-like states are real, the $S$-like states are purely imaginary and that $\Theta |\!\!\uparrow\rangle=+|\!\!\downarrow\rangle$ and $\Theta |\!\downarrow\rangle=-|\!\uparrow\rangle$, i.e. 
\begin{subequations}
\label{eq:lattice-periodic-func-under-TR} 
\begin{align}
\Theta|\Gamma_6,\pm1/2\rangle&=\mp|\Gamma_6,\mp1/2\rangle,\\
\Theta|\Gamma_8,\pm1/2\rangle&=\pm|\Gamma_8,\mp1/2\rangle,\\
\Theta|\Gamma_8,\pm3/2\rangle&=\mp|\Gamma_8,\mp3/2\rangle.
\end{align}
\end{subequations}
Therefore, we can now evaluate e.g.~$\Theta|E+\rangle$ by using Eq.(\ref{eq:lattice-periodic-func-under-TR}) and that $f_{E\Gamma_6}$ is real and $f_{E\Gamma_8}$ is purely imaginary, which gives $\Theta|E+\rangle=-|E-\rangle$. Hence, our conventions lead to
\begin{subequations}
\label{eq:TR-basis-states}
\begin{align}
\Theta|E\pm\rangle&=\mp|E\mp\rangle,\\ 
\Theta|H\pm\rangle&=\mp|H\mp\rangle,
\end{align}
\end{subequations}
which fulfill $\Theta^2=-1$ as expected. We remark that that Rothe \emph{et al.}\cite{Hankiewicz-NJP-2010} find opposite signs under time-reversal (i.e. $\Theta|E\pm\rangle=\pm|E\mp\rangle$ and $\Theta|H\pm\rangle=\pm|H\mp\rangle$), simply because an opposite overall sign was chosen in the definition of  the time-reversal operator.\cite{footnote-Rothe-private}  
In Ref.[\onlinecite{Michetti-Recher-2011}], the same signs as in Eq.(\ref{eq:TR-basis-states}) are found.

\section{Normalization of the BHZ states}\label{App:normalization}

In this Appendix, the normalization of the envelope functions and lattice periodic functions within the envelope function approximation is discussed. To this end, we use $\varphi_{\kk,H+}(\rr)$ in Eq.(\ref{eq:basis-for-BHZ-2D-H+}) as an example. The entire wavefunction is normalized in the usual way, i.e.
\begin{align}
\int_{\mathcal{V}} d \rr |\varphi_{\kk,H+}(\rr)|^2=1, 
\label{eq:normalization-entire-wavefunc}
\end{align}
where $\mathcal{V}$ is the volume of the entire system. The normalization of the entire wavefunction (\ref{eq:normalization-entire-wavefunc}) leaves a freedom to normalize the envelope function and the lattice periodic function in the most convenient way for the problem at hand. Various choices are found in the literature, see e.g.~footnote 2 in the review of Coish and Baugh [\onlinecite{Coish-Baugh-review-2009}]. 

To see how this normalization choice works in practice, we begin by separating the left-hand side of the normalization condition (\ref{eq:normalization-entire-wavefunc}) into a product of the envelope function and the lattice periodic function normalization, respectively. To this end, the normalization condition (\ref{eq:normalization-entire-wavefunc}) is rewriting by dividing the integral over the entire space into a sum of integrals over the unit cells as in Eq.(\ref{eq:int-over-space-to-sum+int-over-unit-cell}), i.e.
\begin{align}
1
&=
\frac{v_a}{L_xL_y} 
\int_{\mathcal{V}}d \rr 
|f_H(z)|^2 
|u_{\Gamma_8,+\frac{3}{2}}(\rr)|^2\nonumber\\
&=\frac{v_a}{L_xL_y} 
\sum_{\Ruc_n}\! 
\int_{v_{uc}^{(n)}}\!\!d \bfrho
|f_H(\rho_z+\mathcal{Z}_n)|^2 
|u_{\Gamma_8,+\frac{3}{2}}(\Ruc_n+\bfrho)|^2\nonumber\\
&\simeq
\frac{v_a}{L_xL_y} 
\sum_{\Ruc_n}
\int_{v_{uc}^{(n)}}d \bfrho
|f_H(\mathcal{Z}_n)|^2 
|u_{\Gamma_8,+\frac{3}{2}}(\bfrho)|^2\nonumber\\
&=
\frac{v_a}{L_xL_y} 
\left[\sum_{\Ruc_n} |f_H(\mathcal{Z}_n)|^2\right]\!
\left[\int_{v_{uc}^{}}\!\!d \bfrho
|u_{\Gamma_8,+\frac{3}{2}}(\bfrho)|^2\right], 
\label{eq:normalization-mellem1}
\end{align}
where we used in the third equality that the envelope function -- by construction -- is slowly varying on the scale of the unit cell, so $f_H(\rho_z+\mathcal{Z}_n)\simeq f_H(\mathcal{Z}_n)$, and that the lattice periodic functions are periodic with the lattice, i.e.~$u_{\Gamma_8,+\frac{3}{2}}(\Ruc_n+\bfrho)=u_{\Gamma_8,+\frac{3}{2}}(\bfrho)$ for all lattice vectors $\Ruc_n$. The integral of $|u_{\Gamma_8,+\frac{3}{2}}(\bfrho)|^2$ over the $n^{\textrm{th}}$ unit cell is the same for every unit cell and hence independent of $n$, which we indicate by $v_{uc}^{(n)}\rightarrow v_{uc}^{}$. Moreover, the sum over lattice points in Eq.(\ref{eq:normalization-mellem1}) can be made into an integral (including the unit cell volume $v_{uc}$), since the envelope function varies slowly on the inter-atomic scale, i.e. 
$\sum_{\Ruc_n} |f_H(\mathcal{Z}_n)|^2
\simeq
\frac{1}{v_{uc}}
\int_{\mathcal{V}}d\Ruc |f_H(\mathcal{Z})|^2$. 
Therefore, we arrive at
\begin{align}
1\!=\!
\frac{v_a}{v_{uc}}\! 
\bigg[\frac{1}{L_xL_y}\!\int_{\mathcal{V}}\!\!d\Ruc |f_H(\mathcal{Z})|^2\bigg]\!
\bigg[\!\int_{v_{uc}^{}}\!\!\!\!\!\!d \bfrho
|u_{\Gamma_8,+\frac{3}{2}}(\bfrho)|^2\bigg], 
\label{eq:normalization-mellem2}
\end{align}
where the normalization of the entire wavefunction in Eq.(\ref{eq:normalization-entire-wavefunc}) have been written as a product of the normalization of the envelope function and lattice periodic function part, respectively. Thus, it is now clear that some freedom exists in the normalization choice. 

In this paper, we normalize the lattice periodic function as Fischer \emph{et al.},\cite{Fischer-Loss-PRL-2010,Fischer-Trauzettel-Loss-PRB-2009,Fischer-Trif-Coish-Loss-Solid-state-comm-2009,Fischer-Coish-Bulaev-Loss-PRB-2008} i.e. 
\begin{align}
\int_{v_{uc}^{}}\!\!d \bfrho
|u_{\Gamma_8,+\frac{3}{2}}(\bfrho)|^2=\frac{v_{uc}}{v_a}=2
\label{eq:lattice-perio-func-normalization}
\end{align}
using that a zinc-blende crystal like HgTe or GaAs contains two atoms per unit cell, $v_{uc}=2v_a$. This normalization has the advantage that the atomic HF constants found in the main paper are independent of the number of atoms in the unit cell.\cite{Coish-Baugh-review-2009} Moreover, the envelope function is normalized as
\begin{align}
\frac{1}{L_xL_y}\int_{\mathcal{V}}d\Ruc |f_H(\mathcal{Z})|^2=1, 
\end{align}
such that Eq.(\ref{eq:normalization-mellem2}) is fulfilled. 

The normalization procedure follows the same lines as above for the other BHZ basis functions, e.g. all lattice periodic functions are normalized to the number of atoms in the unit cell. When the wave function is not a simple product of an envelope function and a lattice periodic function, then it should be used that different lattice periodic functions are orthogonal, i.e. $\int_{v_{uc}} d\bfrho u^\ast_{\Gamma_a,m_j}(\bfrho)u^{}_{\Gamma_b,m'_j}(\bfrho)= 2\delta_{\Gamma_a,\Gamma_b}\delta_{m_j,m'_j}$. Finally, it should be noted that for the $E\pm$ states, we end up with a combined normalization for the two envelope functions, i.e. $\int d z  \big[|f_{E\Gamma_6}(z)|^2+|f_{E\Gamma_8}(z)|^2\big]=1$. We assume that each of these two envelope functions are normalized to one half.\cite{Bernevig-Zhang-Science-2006}   

\section{Details on the calculation of the atomic integrals of the HF interactions for $P$-like states}\label{App:atomic-integrals}

This Appendix deals with the integrals over the atomic wave functions of the form
\begin{align}
\int_{v_{uc}} \!\! d \bfrho
\big[\Psi^{\Hg/\Te}_{\Gamma_8,m_j}(\bfrho\mp\bfd/2)\big]^\ast
h^n_{i}
\Psi^{\Hg/\Te}_{\Gamma_8,m'_j}(\bfrho\mp\bfd/2),
\label{eq:int-over-atomic-orbitals-for-h23-Appendix}  
\end{align}
which appear in the matrix elements of $H_{\textrm{HF},2}$ and $H_{\textrm{HF},3}$ in Sec.~\ref{subsec:HF-for-P-states}. The atomic wave functions are written as $\Psi_{\Gamma_8,m_j}^{\Hg/\Te}(\rr)=R^{\Hg/\Te}(r)\mathbb{Y}_{\Gamma_8,m_j}(\theta,\phi)$, i.e.~a product of a radial and an angular part as in the main text. The angular part of the wave functions $\mathbb{Y}_{\Gamma_8,m_j}(\theta,\phi)$ are  combinations of the usual spherical harmonics\cite{Sakurai-modern-BOOK} $Y_{l}^m(\theta,\phi)$ and the electronic spin-$1/2$ ($|\!\uparrow\rangle$ and $|\!\downarrow\rangle$)  and inherit the symmetry of the bands,\cite{Gueron-PR-1964,Fischer-Coish-Bulaev-Loss-PRB-2008} i.e. 
\begin{align}
\label{eq:angular-parts-p-states}
\mathbb{Y}_{\Gamma_8,+\frac{3}{2}}(\theta,\phi)
&=
-Y_{1}^{1}(\theta,\phi)|\!\uparrow\rangle,\\ 
\mathbb{Y}_{\Gamma_8,+\frac{1}{2}}(\theta,\phi)
&=
-\sqrt{\frac{2}{3}}Y_{1}^{0}(\theta,\phi)|\!\uparrow\rangle
-\sqrt{\frac{1}{3}}Y_{1}^{1}(\theta,\phi)|\!\downarrow\rangle,
\nonumber\\ 
\mathbb{Y}_{\Gamma_8,-\frac{1}{2}}(\theta,\phi)
&=
-\sqrt{\frac{2}{3}}Y_{1}^{0}(\theta,\phi)|\!\!\downarrow\rangle
-\sqrt{\frac{1}{3}}Y_{1}^{-1}(\theta,\phi)|\!\uparrow\rangle,
\nonumber\\ 
\mathbb{Y}_{\Gamma_8,-\frac{3}{2}}(\theta,\phi)
&=
-Y_{1}^{-1}(\theta,\phi)|\!\downarrow\rangle,
\nonumber
\end{align}
which are all eigenfunctions of $J_z=L_z+S_z$ (with eigenvalue $\h m_j$), $\Jj^2=(\Ll+\Ss)^2$ (with $j=3/2$), $\Ll^2$ (with $l=1$ due to $P$-states) and $\Ss^2$ (with $s=1/2$). To be consistent with the BHZ model, we use the same overall sign as Refs.~\onlinecite{Bernevig-Zhang-Science-2006,Novik-Molenkamp-PRB-2005}, which is opposite to the one used in e.g. Refs.~\onlinecite{Winkler-BOOK-2003,Katsaros-Glazman-et-al-PRL-2011} (see also Appendix \ref{App:TR-af-basis-states}, Eq.(\ref{eq:Gamma-states-P}) and endnote \onlinecite{footnote-overall-sign-of-P-states}). However, this overall sign cancels out in the matrix elements between $P$ states and therefore has no effect here. 

To find the integrals (\ref{eq:int-over-atomic-orbitals-for-h23-Appendix}), the spherical approximation Eq.(\ref{eq:speherical-approx-for-unit-cell-int-over-h2-ogh3}) is used. This is an excellent approximation, since most of the weight of the integrals are close to the atomic core. To facilitate the calculations, the HF dipole-dipole like interaction for a single nuclear spin $h_2^n$ Eq.(\ref{eq:HF2-of-nuclei-nr-n}) is rewritten as (choosing the origin at the nuclear spin, i.e.~$\rr_n=\rr-\RR_n\rightarrow \rr$)
\begin{align}
h^{n}_2=&
\frac{\mu_0}{4\pi}\gamma_e\gamma_{j_n}
\frac{1}{r^3(1+\frac{r_c}{r})}\frac{1}{2}
\label{eq:omskrivning-af-h2}\\
&\!\!\times\!\!\Bigg\{\!
\left[\frac{3z^2-r^2}{r^2}\right]
\Big[2S_zI_{z,n}-\frac{1}{2}\big(S_+I_{-,n}+S_-I_{+,n}\big)\Big]
\nonumber\\
&\hspace{3mm}+
3
\left[\frac{x^2-y^2}{r^2}\right]
\frac{1}{2}\big(S_+I_{+,n}+S_-I_{-,n}\big)
\nonumber\\
&\hspace{3mm}+
\frac{6xy}{r^2}(S_xI_{y,n}+S_yI_{x,n})
+\frac{6xz}{r^2}(S_xI_{z,n}+S_zI_{x,n})
\nonumber\\
&\hspace{3mm}+
\frac{6yz}{r^2}(S_yI_{z,n}+S_zI_{y,n})
\Bigg\}.
\nonumber
\end{align}
This is written in such a way that the integrals (like Eq.(\ref{eq:speherical-approx-for-unit-cell-int-over-h2-ogh3})) consist of a radial integral over $\propto\frac{1}{r^3(1+\frac{r_c}{r})}$ times a sum of angular integrals. The terms in the curly bracket become the sum of angular integrals, where the space dependencies are seen to form spherical tensor operators or sums thereof. Therefore, the Wigner-Eckart theorem is useful to identify the integrals that are zero, see e.g.~Ref.[\onlinecite{Sakurai-modern-BOOK}]. As an example, the element for a Hg nuclear spin between the $\Psi_{\Gamma_8,\tau3/2}^{\Hg}(\rr)$ states, appearing in the matrix element $\langle\varphi_{\kk H\tau}|H_{\textrm{HF},2}|\varphi_{\kk' H\tau'}\rangle$, is found to be (after some calculations)
\begin{align}
&\int_0^{r_{\textrm{max}}}\!\!\!\!\!\! d r r^2
\!\!\int_0^{2\pi} \!\!\!\!d \phi 
\!\!\int_0^{\pi}\!\!\! d \theta \sin(\theta)  
[\Psi_{\Gamma_8,\tau3/2}^{\Hg}(\rr)]^\ast
h^n_{2}
\Psi_{\Gamma_8,\tau'3/2}^{\Hg}(\rr)
\nonumber\\
&=
\frac{\mu_0}{4\pi}\gamma_e\gamma_{j_n}
\left\langle\frac{1}{r^3}\right\rangle^{\Hg}_{r}
\delta_{\tau,\tau'}
\Big[
-\frac{1}{5}\tau\h I_{z,n}
\Big]
\end{align}
using Eqs.(\ref{eq:angular-parts-p-states},\ref{eq:omskrivning-af-h2}) and the definition (\ref{eq:def-<1/r^3>}). The rest of the integrals for $h_2^n$ are found similarly. 

Finally, we note that the integrals (\ref{eq:int-over-atomic-orbitals-for-h23-Appendix}) involving $h_3^n$ (\ref{eq:HF3-of-nuclei-nr-n}) are much simpler to evaluate. The integrals in the spherical approximation (\ref{eq:speherical-approx-for-unit-cell-int-over-h2-ogh3}) again separate into the radial integral $\langle1/r^3\rangle^{\Hg/\Te}_{r}$ times a sum of angular integrals, which can be found by using the rewriting  $\mathbf{L}_n\cdot\II_n=L_{z,n}I_{z,n}+\frac{1}{2}[L_{+,n}I_{-,n}+L_{-,n}I_{+,n}]$ and Eq.(\ref{eq:angular-parts-p-states}).

\section{Estimation of the Atomic HF constants}\label{App:estimation-of-HF-constants}  

In this Appendix, we estimate the atomic HF couplings Eqs.(\ref{eq:atomic-HF-couplings-def},\ref{eq:atomic-P-HF-coupling-def}),   
\begin{subequations}
\begin{align}
A^{\textrm{Atomic}}_{S,j_n}
&=
\frac{2\mu_0}{3} 
g_e\mu_Bg_{j_n}\mu_N
|u_{\Gamma_6}(\RR_n)|^2,\\
A^{\textrm{Atomic}}_{P,j_n}
&=
\frac{\mu_0}{4\pi}
g_e\mu_Bg_{j_n}\mu_N
(N_{\Gamma_8})^2|\alpha_{j_n}|^2
\left\langle\frac{1}{r^3}\right\rangle^{j_n}_{r} 
\end{align}
\end{subequations}
along the same lines as Fischer \emph{et al.}\cite{Fischer-Coish-Bulaev-Loss-PRB-2008} These estimates are given in table \ref{Table:atomic-S-HF-couplings}. Below, we go through the ingredients to make these estimates.

Within the LCAO approach (\ref{eq:u6-og-u8-in-the-LCAO-approach}), the lattice periodic functions within a unit cell are written as a linear combination of the two atomic orbitals. The relative weight between the two orbitals is related to the ionicity and found to be\cite{Willig-Sapoval-1977,footnote-uncertainty-in-ionicity} 
\begin{align}
\alpha_{\Te}\simeq \sqrt{0.8}  
\qquad\textrm{and}\qquad
\alpha_{\Hg}\simeq \sqrt{0.2}, 
\end{align}
which is taken to be the same for the $\Gamma_6$ and $\Gamma_8$ bands.\cite{Fischer-Coish-Bulaev-Loss-PRB-2008} 

Moreover, the $g$-factors for the various isotopes are\cite{Schliemann-Khaetskii-Loss-JoPCM-Review-2003}
\begin{subequations}
\begin{align}
g_{^{199}\textrm{Hg}}^{}&=1.01,
\quad
g_{^{201}\textrm{Hg}}^{}=-0.37,\\
g_{^{123}\textrm{Te}}^{}&=-1.47,
\quad
g_{^{125}\textrm{Te}}^{}=-1.78.
\end{align}
\end{subequations}
These are seen to vary in sign, which is the reason for the sign variation of the HF couplings.   

Furthermore, to estimate the HF couplings, the atomic wave functions $\Psi_{\Gamma_i,m_j}^{\Hg/\Te}$ also have to be given explicitly. The angular part follow the band symmetry as in Eq.(\ref{eq:angular-parts-p-states}). As for the radial part, we follow Fischer \emph{et al.}\cite{Fischer-Coish-Bulaev-Loss-PRB-2008} and approximate it by a hydrogenic radial eigenfunction\cite{Sakurai-modern-BOOK} $R_{nl}(r)$ with an effective charge $eZ_{\textrm{eff}}$ replacing the actual charge of the nucleus $eZ$ in order to include atomic screening effects etc., i.e.~$Z_{\textrm{eff}}<Z$. The outermost electrons in Hg (Te) have the principal quantum number $n=6$ ($n=5$) such that $R^{\Hg(\Te)}_{\Gamma_8}(r)=R_{6(5),1}(r)$ and  $R^{\Hg(\Te)}_{\Gamma_6}(r)=R_{6(5),0}(r)$. Clementi \emph{et al.}\cite{Clementi-Raimondi-1963,Clementi-Raimondi-Reinhardt-1967} have calculated the effective charges $Z_{\textrm{eff}}$ for various atoms and orbitals and found that $Z_{\textrm{eff}}(\textrm{Te},5s)=12.5$, $Z_{\textrm{eff}}(\textrm{Te},5p)=10.8$ and $Z_{\textrm{eff}}(\textrm{Hg},6s/6p)=11.2$, which obviously is much smaller than the bare nuclear charges  $eZ=52e$ for Te and $eZ=80e$ for Hg.  

Using $|u_{\Gamma_6}(\RR_n)|^2\simeq(N_{\Gamma_6,1/2})^2|\alpha^{}_{j}|^2|\Psi^{j}_{\Gamma_6}(\mathbf{0})|^2$ with the hydrogenic orbital $\Psi^{j}_{\Gamma_6}(\rr)=R_{n0}(r)Y_0^0(\theta,\phi)$ for isotope $j$, we can now give the atomic contact HF coupling as
\begin{align}
A^{\textrm{Atomic}}_{S,j}
\simeq&
\frac{2\mu_0}{3} 
g_e\mu_Bg_{j}\mu_N
[N_{\Gamma_6,1/2}]^2|\alpha^{}_{j}|^2
\frac{[Z_{\textrm{eff}}(j,s)]^3}{\pi a_0^3n^3},
\label{eq:S-HF-coupling-appendix}
\end{align}
where $a_0=4\pi \epsilon_0\h^2/(m_ee^2)$ is the Bohr radius and $\epsilon_0$ is permittivity of free space. 

The hydrogenic like atomic orbitals also makes it easy to calculate $\langle1/r^3\rangle^{j}_{r}$ in Eq.(\ref{eq:def-<1/r^3>}) numerically, which shows that neither the nuclear length scale $r_c$ nor $r_{\textrm{max}}$ make a difference in practice. Hence, we can use  
\begin{align}
\left\langle\frac{1}{r^3}\right\rangle^{j}_{r}
\simeq
\int_0^{\infty}\!\!\!d r  r^2
|R_{n,l=1}(r)|^2
\frac{1}{r^3}
=
\frac{[Z_{\textrm{eff}}(j,p)]^3}{3 a_0^3n^3},
\label{eq:P-HF-coupling-appendix}
\end{align}
to find the HF coupling $A^{\textrm{Atomic}}_{P,j}$ for the $P$ states.

Therefore, now we only need  one more ingredient to be able to estimate the HF couplings, namely the normalization constants $N_{\Gamma_i,m_j}$ of the lattice periodic functions in the LCAO approach (\ref{eq:u6-og-u8-in-the-LCAO-approach}). The normalization condition (\ref{eq:lattice-perio-func-normalization}) lead to
\begin{align}
&
\int_{v_{uc}}\!\!\!\!\!\! d\rr 
\Big|\alpha^{}_{\Te}\Psi^{\Te}_{\Gamma_i,m_j}\Big(\!\rr{+}\frac{\bfd}{2}\Big)
{\pm}\alpha^{}_{\Hg}\Psi_{\Gamma_i,m_j}^{\Hg}\Big(\!\rr{-}\frac{\bfd}{2}\Big)\Big|^2
\!\!=\!
\frac{2}{N_{\Gamma_i,m_j}^2},
\nonumber
\end{align}
where $+$ ($-$) corresponds to $\Gamma_8$ ($\Gamma_6$). First of all, we note that 
\begin{align}
N_{\Gamma_i,m_j}=N_{\Gamma_i,-m_j}
\label{eq:normalization-consts-uaf-af-fortegn-af-mj} 
\end{align}
due to the similar form of the atomic wave functions for $\pm m_j$, see e.g. Eq.(\ref{eq:angular-parts-p-states}). Using the hydrogenic eigenstates, we can therefore now numerically find the normalization constants $N_{\Gamma_i,m_j}$. Numerically, these do depend weakly on how the Wigner-Seitz unit cell of the zinc-blende crystal is approximated, in contrast to the unit cell integrals involving $h_{i}^n$ for $i=2,3$ in Sec.~\ref{subsec:HF-for-P-states}. We have tested various spherical and cubic approximations to the primitive Wigner-Seitz unit cell all with the same volume as the Wigner-Seitz unit cell, namely $v_{uc}=16|\bfd|^3/(3\sqrt{3})$, where $|\bfd|=0.279$nm is the distance between the Hg and Te atoms in unit cell, see e.g.~p.58 in Ref.[\onlinecite{Stoneham-BOOK}]. Such a weak dependence is also found in the estimate for GaAs by Fischer \emph{et al}.\cite{Fischer-Coish-Bulaev-Loss-PRB-2008} From our various approximate unit cell calculation, we found that a good estimate for the normalization constants are $(N_{\Gamma_8,1/2})^2\simeq (N_{\Gamma_8,3/2})^2\simeq 3.6$ and $(N_{\Gamma_6,1/2})^2\simeq2.7$. Therefore, we can use approximately equal normalization constants for $u_{\Gamma_8,\pm1/2}$ and $u_{\Gamma_8,\pm3/2}$, which allows for the introduction of a common 	atomic $P$-like HF constant in Eq.(\ref{eq:atomic-P-HF-coupling-def}). Therefore, we now have all the ingredients to make the estimates with the results seen in table \ref{Table:atomic-S-HF-couplings}. 
 
\section{Hyperfine interactions in terms of edge state spin operators}\label{app:HF-in-spin-operators}

Here we reformulate the HF interactions (\ref{eq:HF1-HES-basis-along-y-axis-general}) and (\ref{eq:H-HF23-ES-basis-general-form}) for the HESs along the $y$-axis in order to give some more insights into their form. Having in mind the spin-1/2 picture of a pair of HESs discussed in sec.~\ref{sec:HES}, we are lead to introduce the \emph{non-diagonal edge states spin operators} as
\begin{subequations}
\label{eq:HES-spin-operators-def}
\begin{align}
\mathfrak{s}_{k_yk_y',x}
&=\frac{\h}{2}\Big[c_{k_yu}^\dag c_{k_y'd}^{} +c_{k_yd}^\dag c_{k_y'u}^{} \Big],
\\
\mathfrak{s}_{k_yk_y',y}
&=i\frac{\h}{2}
\Big[c_{k_yd}^\dag c_{k_y'u}^{} -c_{k_yu}^\dag c_{k_y'd}^{} \Big],
\\
\mathfrak{s}_{k_yk_y',z}
&=\frac{\h}{2}\Big[c_{k_yu}^\dag c_{k_y'u}^{} -c_{k_yd}^\dag c_{k_y'd}^{} \Big],
\end{align}
\end{subequations}
together with the operator $\mathfrak{I}_{k_yk_y'}=\frac{\h}{2}[c_{k_yu}^\dag c_{k_y'u}^{} +c_{k_yd}^\dag c_{k_y'd}^{} ]$ and the raising and lowering operators $\mathfrak{s}_{k_yk_y',\pm}\equiv \mathfrak{s}_{k_yk_y',x}\pm i\mathfrak{s}_{k_yk_y',y}$ for the edge state spin. Here, for instance,  $\mathfrak{s}_{k_yk_y',+}$ moves a particle in the state $\Psi^{d}_{y,k_y'}$ into the state $\Psi^{u}_{y,k_y}$ and in this sense raises the edge state spin (while also changing the wave vector). In the case of $k_y=k_y'$, the edge state spin operators (\ref{eq:HES-spin-operators-def}) coincide with the usual spin-1/2 operators\cite{Flensberg-BOOK} and $\mathfrak{I}_{k_yk_y}$ is the particle number operator  (times $\h/2$).    

The contact HF interaction (\ref{eq:HF1-HES-basis-along-y-axis-general}) in terms of the edge state spin operators (\ref{eq:HES-spin-operators-def}) becomes
\begin{align}
\mathcal{H}^{(y)}_{\textrm{HF},1}
=& 
\frac{1}{2\h^2}
\frac{B-D}{2B} 
\sum_{k_yk_y'}
\sum_n
\frac{e^{i(k_y'-k_y)\mathcal{Y}_n}}{L_y}
A_{S,j_n}(\mathcal{Z}_n)
\\
&\times
\bigg[
\big(
\Lambda^{(\mathcal{X}_n)}_{k_y,k_y'}
+
\Lambda^{(\mathcal{X}_n)}_{-k_y,-k_y'}
\big)
I_{z,n}
\mathfrak{s}_{k_yk_y',z}
\nonumber\\
&\hspace{4mm}
+
\big(
\Lambda^{(\mathcal{X}_n)}_{k_y,k_y'}
-
\Lambda^{(\mathcal{X}_n)}_{-k_y,-k_y'}
\big)
I_{z,n}
\mathfrak{I}_{k_yk_y'}
\nonumber\\
&\hspace{2mm}
-
\Lambda^{(\mathcal{X}_n)}_{k_y,-k_y'}
I_{-,n}
\mathfrak{s}_{k_yk_y',+}
-
\Lambda^{(\mathcal{X}_n)}_{-k_y,k_y'}
I_{+,n}
\mathfrak{s}_{k_yk_y',-}
\bigg].
\nonumber
\end{align}
Using the edge state spin operators (\ref{eq:HES-spin-operators-def}), the HF interaction (\ref{eq:H-HF23-ES-basis-general-form}) due to the $P$-like states becomes
\begin{align}
&\mathcal{H}^{(y)}_{\textrm{HF},P}
=
\frac{2}{15\h^2}
\sum_n 
\sum_{k_y,k_y'}
\frac{e^{i(k_y'-k_y)\mathcal{Y}_n}}{L_y}
\Bigg\{
-2\frac{B-D}{B}
\nonumber\\
&\times
A_{P,j_n}^{EE}
\Big[
\Lambda^{(\mathcal{X}_n)}_{k_y,-k_y'}
I_{-,n}
\mathfrak{s}_{k_yk_y',+}
+
\Lambda^{(\mathcal{X}_n)}_{-k_y,k_y'}
I_{+,n}
\mathfrak{s}_{k_yk_y',-}
\Big]
\nonumber\\
&\hspace{7mm}
+
\big(
\Lambda^{(\mathcal{X}_n)}_{k_y,k_y'}
+\Lambda^{(\mathcal{X}_n)}_{-k_y,-k_y'}
\big)
\big[
\mathfrak{L}_nI_{x,n}
+
\mathfrak{F}_nI_{z,n}
\big]
\mathfrak{s}_{k_yk_y',z}
\nonumber\\
&
\hspace{7mm}
+
\big(
\Lambda^{(\mathcal{X}_n)}_{k_y,k_y'}
-\Lambda^{(\mathcal{X}_n)}_{-k_y,-k_y'}
\big)
\big[
\mathfrak{L}_nI_{x,n}
+
\mathfrak{F}_nI_{z,n}
\big]
\mathfrak{I}_{k_yk_y'}
\Bigg\},
\nonumber
\end{align}
where
\begin{align} 
\mathfrak{L}_n&\equiv-\frac{\sqrt{3}A\sqrt{B^2-D^2}}{|A|B}2\textrm{Im}(A_{P,j_n}^{EH}),\nonumber\\ 
\mathfrak{F}_n&\equiv
\frac{1}{B}
\big[
(B-D)A_{P,j_n}^{EE}+3(B+D)A_{P,j_n}^{HH}
\big].
\end{align}
In both HF interactions the edge state spin-flipping terms $I_{\pm,n}\mathfrak{s}_{k_yk_y',\mp}$ appear. Moreover, in the HF interaction for $P$-like states the unusual coupling $I_{x,n}\mathfrak{s}_{k_yk_y',z}$ is found as discussed in the main text. Note that the terms including the operator  $\mathfrak{I}_{k_yk_y'}$ vanish to second order in $k_y$ and $k_y'$ in the position averaging and also in the particle-hole symmetric limit, see sec.~\ref{sec:position-average}.


\begin{thebibliography}{124}
\bibitem[{foo({\natexlab{b}})}]{footnote-r_c-cut-off}
\bibinfo{note}{Many authors simply neglect $r_c$ in $h_i^n$ ($i=2,3$), since
  the only role it plays, is to see explicitly that the dipole-dipole like
  interaction $h_2^n$ is zero for an $S$-like state. If $r_c$ is absent, then
  the matrix element of the dipole-dipole like interaction is not well-defined,
  see e.g. p.455 in [\onlinecite{Stoneham-BOOK}]. However, since $r_c$ is so
  small, no non-zero matrix element depends on its magnitude.}
\bibitem[{foo({\natexlab{c}})}]{footnote-zero-energy}
\bibinfo{note}{We have chosen the zero of energy such that $\e_k$ is zero at
  $\kk=0$.}
\bibitem[{foo({\natexlab{d}})}]{footnote-lattice-periodic-func}
\bibinfo{note}{The lattice periodic functions\cite{Ashcroft-Mermin-BOOK}
  $\langle \rr|\Gamma_i, m_j,\kk\rangle\equiv u_{\kk,\Gamma_i, m_j}(\rr)$ enter
  the Bloch states as $\psi_{\kk,\Gamma_i,
  m_j}(\rr)=e^{i\kk\cdot\rr}u_{\kk,\Gamma_i, m_j}(\rr)$ and have the
  periodicity of the lattice, i.e. $u_{\kk,\Gamma_i,
  m_j}(\rr+\Ruc_n)=u_{\kk,\Gamma_i, m_j}(\rr)$ for all Bravais lattice vectors
  $\Ruc_n$.}
\bibitem[{foo({\natexlab{e}})}]{footnote-atomic-vol}
\bibinfo{note}{In the case of a zincblende crystal like HgTe or GaAs, the
  atomic volume is half of the two-atomic primitive unit cell volume in real
  space.}
\bibitem[{foo({\natexlab{i}})}]{footnote-unit-and-k-dependence-in-HF-constant}
\bibinfo{note}{Note that we keep the factor
  $e^{i(\kk'-\kk)\cdot\Ruc_{n\bot}^{}}/L_xL_y$ outside the definition of the
  position dependent HF coupling in order to have $A^{S}_{j_n}(\mathcal{Z}_n)$
  independent of $\kk$. However, this choice means that the unit of
  $A^{S}_{j_n}(\mathcal{Z}_n)$ is not energy, but energy times length squared.}
\bibitem[{foo({\natexlab{j}})}]{footnote-prefactor-HF1}
\bibinfo{note}{The prefactor of $1/2$ is the spin projection of the electron on
  the $z$-axis and therefore not included in $A^{S}_{j_n}(\mathcal{Z}_n)$ in
  order to have the Hamiltonian in a form analogue to the quantum dot case,
  where $H=\sum_nA_n\Ss\cdot\II_n$.}
\bibitem[{foo({\natexlab{k}})}]{footnote-atomic-parity}
\bibinfo{note}{The parity $\mathcal{P}$ around the atomic core for the
  atomic-like wave functions $\Psi_{\Gamma_6,m_j}^{\Hg/\Te}$ is even
  ($S$-like), whereas $\Psi_{\Gamma_8,m_j}^{\Hg/\Te}$ has odd parity
  ($P$-like). This follows e.g. from the parity of the spherical harmonics.
  Since the dipole-dipole interaction is invariant under atomic parity,
  $\mathcal{P}h_2^n\mathcal{P}^{-1}=+h_2^n$, the matrix elements between $S$-
  and $P$-like states vanish.}
\bibitem[{foo({\natexlab{l}})}]{footnote-chosing-r_max}
\bibinfo{note}{For instance, we could choose $r_{\textrm{max}}$ such that half
  of the unit cell volume is in a sphere around the Hg atom, i.e.
  $v_{uc}/2=4\pi r_{\textrm{max}}^3/3$, or $r_{\textrm{max}}$ could be chosen
  to be equal to half the distance between the atoms $|\bfd|/2$. However, in
  either case $r_{\textrm{max}}$ is larger then $a_0$ such that the numerical
  value of the integral do not change -- even if $r_{\textrm{max}}$ is set to
  infinity.}
\bibitem[{foo({\natexlab{m}})}]{footnote-different-P-HF-couplings-possible}
\bibinfo{note}{In principle, an atomic HF coupling could be introduced for
  $h_2^n$ and $h^n_3$ separately. The only difference between the two would be
  their normalization constants $N_{\Gamma_8,m_j}$ with
  $m_j=\frac{3}{2},\frac{1}{2}$, respectively. However, due to their numerical
  similarity, $N_{\Gamma_8,3/2}\simeq N_{\Gamma_8,1/2}$, we consider this to be
  an unnecessary complication in the case of HgTe.}
\bibitem[{foo({\natexlab{o}})}]{footnote-normalization-general}
\bibinfo{note}{The normalization of the envelope functions is $\int
  d\rr_\bot\sum_{\zeta=E\pm,H\pm} |\phi^{}_{\eta,\zeta}(\rr_{\bot})|^2=1$,
  which can be showed along the lines of Appendix \ref{App:normalization}.}
\bibitem[{foo({\natexlab{r}})}]{footnote-estimate-number-of-atoms-within-HES}
\bibinfo{note}{This estimate is based on a 7nm thick quantum well with a HES
  width of about $W_x\sim\lambda_2^{-1}\simeq 50$nm for $k=0$ and using that
  the atomic volume $v_a$ is half of the Wigner-Seitz unit cell volume, i.e.
  $v_a=8|\bfd|^3/(3\sqrt{3})\simeq 3.3\times 10^{-2}\textrm{nm}^3$ for the
  zincblende crystal, where $|\bfd|=0.279$nm is the bond length between the Hg
  and Te atoms.\cite{Stoneham-BOOK}}.
\end{thebibliography}

\end{document}